\documentclass[12pt]{article}

\usepackage{amssymb,amsmath,graphicx}

\begin{document}

\begin{titlepage}

\begin{center}

\vspace{0.8cm}

{\Large Baryogenesis via left-right asymmetry generation 
by Affleck-Dine mechanism
in Dirac neutrino model}

\vspace{1.3cm}

{\bf Masato Senami}\footnote{senami@icrr.u-tokyo.ac.jp}
and
{\bf Tsutomu Takayama}\footnote{tstkym@icrr.u-tokyo.ac.jp}
 \\

\vspace{1cm}

{\it
ICRR, University of Tokyo, Kashiwa 277-8582, Japan
}

\vspace{1.5cm}

\abstract{
A baryogenesis scenario in supersymmetric standard models with Dirac neutrinos
proposed by Abel and Page is reconsidered
with introducing intermediate scale physics
to stabilize the runaway potential along a right-handed sneutrino direction.
In contrary with previous results,
the baryon number asymmetry can be explained even for higher reheating temperature
without entropy production if the lightest neutrino mass is small
and/or thermal effects induce early oscillation.
We discuss the solution to the problem of
dark matter overproduction by the right-handed sneutrino decay
by SU(2)$_R$ gauge interaction.
}

\end{center}
\end{titlepage}

\section{Introduction}

There are convincing evidences for small non-zero neutrino masses.
From solar and atmospheric neutrino oscillation data, 
finite squared-mass differences of neutrinos are reported as \cite{Experiments,Fogli:2005cq}
\begin{eqnarray}
	\label{eq:constraint_finite_neutrino_mass}
	\Delta m_{\rm sol}^2 \simeq 7.9\times10^{-5}{\mathrm{eV}}^2,
	~~~ \Delta m_{\rm atm}^2 \simeq 2.4\times10^{-3}{\mathrm{eV}}^2 .
\end{eqnarray}
On the other hand, an upper bound of neutrino masses is derived by
the observation of the cosmic microwave background (CMB) as
$\sum_i|m_i|<2.0$ eV \cite{Ichikawa:2004zi,Spergel:2006hy}.
In the standard model (SM) the neutrinos have no mass term.
Hence, the SM should be extended to include small non-zero neutrino masses.

The origin of baryon number asymmetry of the universe is another important problem,
which is not explained in the SM.
Recently, the existence of the baryon number asymmetry is confirmed by WMAP 
with considerable accuracy \cite{Spergel:2006hy}.
That is given in terms of baryon-to-entropy ratio as
\begin{eqnarray}
	\label{eq:constraint_baryon-to-entropy_ratio}
	\frac{n_B}{s}= (8.7{\pm 0.3}) \times10^{-11}
\end{eqnarray}
where $s$ is the entropy density.

Introducing heavy right-handed Majorana neutrinos into the SM
solves simultaneously these two problems
by the seesaw mechanism \cite{seesaw} and leptogenesis \cite{Fukugita:1986hr}.
In the seesaw mechanism, the smallness of the neutrino mass is naturally explained
from the suppression by heavy Majorana masses of right-handed neutrinos.
The Majorana mass term for neutrinos does not conserve lepton number, $L$.
Hence, lepton number asymmetry can be generated through this Majorana mass term
\cite{leptogenesis}.
Once non-zero lepton number asymmetry is generated,
it is transferred to baryon number asymmetry
via the $B+L$ violating sphaleron process of the SU(2)$_L$ gauge theory.
Therefore, the SM with heavy right-handed Majorana neutrinos
can explain simultaneously neutrino masses and the baryon number asymmetry of the universe.

The small neutrino mass can be also explained by the Dirac mass
with a very small Yukawa coupling, $y_\nu$.
In the Dirac neutrino model, the smallness of the neutrino mass is replaced by
the smallness of the Yukawa coupling, $y_\nu = m_\nu / v \sim 10^{-13}$,
where $v \sim 100$ GeV is the vacuum expectation value (vev) of the Higgs boson.
If this coupling appears after the some GUT particles are integrated out,
it is expected to be suppressed by $m_W / M_{\rm GUT}$
\cite{diracneutrino}.
Hence, the Dirac neutrino mass is also well motivated
for the smallness of the neutrino mass.
On the other hand, the Dirac neutrino model is not considered to be 
well motivated for leptogenesis,
since this model does not introduce new lepton number violation.

However, the baryon number asymmetry can be generated 
via neutrinogenesis scenarios with Dirac neutrinos 
without introducing new lepton or baryon number violation \cite{Dick:1999je}.
Neutrinogenesis scenario is based on the idea of hiding lepton number in inert species
from the sphaleron process \cite{Campbell:1992jd}.
Following these works, very interesting idea was recently proposed
in supersymmetric (SUSY) extension of the SM with Dirac neutrino \cite{Abel:2006hr}.
In this scenario, left-right asymmetry ($L^{(L)}-L^{(R)} \neq 0$)
is produced through the A-term of the neutrino Yukawa term
by the Affleck-Dine mechanism \cite{Affleck:1984fy,Dine:1995kz}
in the dynamics of the combination of $LH_u$ flat direction and right-handed sneutrino
\cite{manifold}
without generating net lepton number asymmetry ($L^{(L)} + L^{(R)} = 0$).
Then only the left-handed lepton asymmetry ($L^{(L)}$) is transferred to baryon number
asymmetry via the sphaleron process, which acts only on left-handed particles,
 while the right-handed lepton asymmetry
($L^{(R)}$) remains in the right-handed sneutrinos
since right-handed sneutrinos are out of equilibrium 
due to their tiny Yukawa couplings
while the sphaleron process is effective.

Nevertheless, there are a few unclear points in the original work \cite{Abel:2006hr}.
First, the right-handed sneutrino direction is not bounded in the potential
unless we consider non-renormalizable terms.
Hence, the right-handed sneutrino field runs away during the inflation.
Therefore, the initial values of Affleck-Dine scalar fields (AD-fields)
studied in Ref.~\cite{Abel:2006hr} are local minimum and may be unstable.
Second, although the AD-fields were assumed to track the local minimum
determined by the balance between Hubble-indecued negative mass terms and quartic terms,
these AD-fields may begin to oscillate soon after the inflation ends and
Hubble parameter begin to decrease.
Furthermore, there is a dark matter overproduction problem in this scenario.
Right-handed sneutrino is inevitably produced to the amount of 
the same order of the baryon number asymmetry.
With $R$-parity conservation,
right-handed sneutrinos decay into the lightest SUSY particle (LSP)
after the freeze-out of dark matter.
Hence, the right-handed sneutrino produces dark matter
whose amount is almost the same order of the baryon number.
Therefore, the LSP has too large abundance and overcloses the universe
unless the LSP mass is less than $1{\mathrm{GeV}}$.

In this paper, we reconsider this baryogenesis scenario
assuming intermediate scale physics and including thermal effects
\cite{Allahverdi:2000zd,Anisimov:2000wx}.
The former stabilizes the potential and allows global minimum
to be initial condition of the AD-fields.
With including thermal effects, we can trace the evolution of the AD-fields
from early enough to estimate the baryon number asymmetry.
Then, we estimate the baryon number asymmetry depending on
the intermediate scale and the reheating temperature.
We also discuss the dark matter overproduction problem.
This can be solved by the presence of SU(2)$_R$ gauge symmetry,
which is broken in an intermediate scale.
By this interaction,
the right-handed sneutrino can decay before the freeze-out of dark matter.

This paper is organized as follows.
The set up and the potential for the AD-fields are presented in next section.
In Section 3, we consider the evolution of the AD-fields and
production of the left-right asymmetry via the Affleck-Dine mechanism.
In Section 4, we estimate the baryon number asymmetry depending on
intermediate scales and the reheating temperature.
In Section 5, we mention the dark matter overproduction problem.
Finally, Section 6 is devoted to the summary.

\section{Model}


We consider the extension of the minimal SUSY standard model (MSSM)
by including right-handed neutrinos required for Dirac neutrino mass.
The superpotential is given by Yukawa coupling between the left-handed lepton doublet $L$
and the right-handed neutrino $ \nu_R $
besides the MSSM superpotetial 
\begin{eqnarray}
	\label{eq:superpotential}
	W=W_{\mathrm{MSSM}} + y_\nu L{H_u} \nu_R.
\end{eqnarray}
Non-zero neutrino mass is given by the Higgs mechanism.
For the smallness of the neutrino mass,
the Yukawa coupling constant $y_\nu$ should be very small,
$y_\nu = m_\nu / \langle H_u \rangle  \lesssim \mathcal{O}(10^{-12})$,
where $\langle H_u \rangle \sim {\mathcal{O}}(100) {\mathrm{GeV}}$
is the vev of $H_u$.

There are many flat directions in the MSSM \cite{Gherghetta:1995dv}.
Among them, the $L H_u$ flat direction is most extensively studied direction.
It is parameterized with a complex scalar field $\phi$ as
\begin{eqnarray}
	\label{eq:flat-direction}
	\tilde L =\frac{1}{\sqrt{2}}\left(
	\begin{array}{c}	
		\phi \\ 0
	\end{array}
	\right),
	H_u=\frac{1}{\sqrt{2}}\left(
	\begin{array}{c}
		0 \\ \phi
	\end{array}
	\right),
\end{eqnarray}
where $\tilde L$ and $H_u$ indicate the scalar component of each superfields.
In this model, this direction is not exactly $F$-flat due to the presence of
the $F$-term contribution from the neutrino Yukawa coupling,
\begin{eqnarray}
	\label{eq:F-term_potential}
	V_{\rm quar} = \frac{y_\nu^2}{4}|\phi|^4 + y_\nu^2|\phi|^2|\tilde{\nu}_R|^2.
\end{eqnarray}
However, due to the smallness of $y_\nu$ this contribution is
small enough to allow $\phi$ to develop large value.
Hence, this direction is safely considered to be flat.
Moreover, the smallness of $y_\nu$ also allows $\tilde \nu _R$ to be flat direction simultaneously.
Therefore, we consider the evolutions of these two flat directions.

\section{Evolutions of Affleck-Dine fields}

The scalar potential relevant to the AD-fields $ \phi $ and $\tilde \nu _R $ is given by
\begin{eqnarray}
	\label{eq:scalar_potential}
	V(\phi,{\tilde{\nu}}_R)
	&=& m_\phi^2|\phi|^2 + m_{\tilde{\nu}_R}^2|\tilde{\nu}_R|^2
	+ y_\nu ( A m_{3/2} \phi^2\tilde{\nu}_R + {\mathrm{h.c.}})
    \nonumber \\
	&&-c_\phi H^2 |\phi|^2 - c_{\tilde{\nu}_R}H^2|\tilde{\nu}_R|^2 
	+ y_\nu ( a H\phi^2 \tilde{\nu}_R + {\mathrm{h.c.}})
	\nonumber \\
	&& + \frac{y_\nu^2}{4}|\phi|^4 + y_\nu^2|\phi|^2|\tilde{\nu}_R|^2
	\nonumber \\
	&& + \sum_{f_k|\phi|<T} c_kf_k^2T^2|\phi|^2
	+ \sum_{f_k|\phi|>T} a_{\rm th} \alpha_s^2(T)T^4 
	\ln\left(\frac{|\phi|^2}{T^2}\right) .
\end{eqnarray}
The first and second lines represent soft SUSY breaking terms.
The former one comes from the hidden sector SUSY breaking,
while the latter one is due to the non-zero energy in the early universe
parameterized by the Hubble parameter \cite{Dine:1995kz,Dine:1995uk}.
Here $ m_\phi \sim m_{\tilde \nu _R} \sim m_{3/2}$,
and $A,a,c_{\phi (\tilde \nu_R)}$ are $\cal O$(1) coefficients
depending on the detail of a supergravity model.
The third line is the $F$-term potential from the neutrino Yukawa coupling
as mentioned above.

The fourth line shows thermal corrections to the potential.
Since we consider the evolution of the AD-fields after the inflation ends
until the left-right asymmetry is fixed,
there is high-temperature thermal bath, which interacts with $\phi$,
if the decay products of inflaton are thermalized promptly.
Therefore, we should include thermal corrections in order to discuss the evolution appropriately.
The first term is so-called thermal-mass terms \cite{Allahverdi:2000zd},
which represents one-loop corrections 
to the potential from light particles in the thermal bath.
Here, $f_k$ denote coupling constants of interactions
between left-handed leptons or up-type Higgs
and particles interacting with them,
and coefficients $c_k \sim 1$ are determined
by degrees of freedom of these particles.
These two parameters are summarized in Table \ref{tb:coefficients_of_thermal-mass}.
The temperature of the thermal bath $T$ before the reheating takes place
is estimated as
\begin{eqnarray}
	\label{eq:temperature}
	T \sim (H T_R^2 M_{\mathrm{Pl}})^\frac{1}{4}.
\end{eqnarray}
Here, $T_R$ is the reheating temperature after the inflation and
$M_{\rm Pl} \simeq 2.4 \times 10^{18}$ GeV is the reduced Planck mass.
The condition $f_k|\phi| < T$ 
means effective masses of light degrees induced by $\phi$
should be less than the temperature of the thermal bath.
Light degrees of freedom satisfying this condition
can enter the thermal bath.

\begin{table}[tp]
\begin{center}
	\begin{tabular}{c|cccc}
		~ & leptons	& quarks & W-bosons & Z-boson \\
		\hline
		\hline
		$f_k$ & $y_l/\sqrt{2}$ & $y_q/\sqrt{2}$ & $g_2/\sqrt{2}$ & $\sqrt{(g_1^2 + g_2^2)/2}$ \\
		\hline
		$c_k$ & $1/4$ & $3/4$ & $1/2$ & $1/4$
	\end{tabular}
	\caption{\small Coefficients for thermal-mass terms.
	The coupling $y_l,(y_q, q= t,c,u)$ is the Yukawa coupling
	for charged leptons (up-type quarks) and
	$g_1(g_2)$ is the gauge coupling for the U(1)$_Y$(SU(2)$_L$) gauge group. }
	\label{tb:coefficients_of_thermal-mass}
\end{center}
\end{table}

The second term, thermal-log terms, in the fourth line
comes from thermal corrections in two loop.
These terms originate from the modification of the SU(3)$_C$ gauge coupling
by effects of massive particles integrated out \cite{Anisimov:2000wx}.
Since effective masses of MSSM particles depend on $\phi$, 
corrections to the potential of $\phi$ are induced.
Here $ a_{\rm th} $ is estimated as $a_{\rm th} \simeq 0.47 T(R_i)$
where $T(R_i) = 1/ 2$ for the fundamental representation and 
$ \alpha_s $ is the strong coupling constant.

\subsection{Initial condition}

During the inflation, the Hubble parameter takes almost a constant value, $H_{\rm inf}$.
In this era, the AD-fields quickly settle into
one of the minima of the scalar potential, Eq.~(\ref{eq:scalar_potential}).
However, this potential is a runaway potential;
obviously $V\to-\infty$ for $\phi=0$ and $|\tilde{\nu}_R|\to\infty$.
Therefore, we assume that some gauge symmetry broken at $M_I$
stabilizes the potential for $ \tilde \nu _R $.
For example, $D$-term potential appears when $|{\tilde \nu}_R| (|\phi|) >M_I $,
$V_I = g_I^2 (e_\phi |\phi|^2 + e_{\tilde \nu _R}  |\tilde \nu_R|^2)^2 $
dependent on the charge $e_{\tilde{\nu}_R} (e_{\phi} )$ of $\tilde{\nu}_R(\phi)$.
Hence, we can simply assume that AD-fields are fixed at $M_I$,
unless $ g_I $ is extremely small.
The baryon number asymmetry generated in this model
is independent of details of the intermediate physics
and is only dependent on the initial values.
Considering the possibility that
different intermediate physics stabilize $ \phi $ and $ \nu_R $ respectively,
we take the initial condition $|\phi|=M_{I_L}$ and $|\tilde{\nu}_R|=M_{I_R}$.

Other stabilization mechanisms are of course available,
such as non-renormalizable terms.
For those cases, initial conditions, evolutions of the AD-fields and
estimates of the left-right asymmetry are quantitatively changed.
However, our analysis can be extended easily to other stabilization mechanism.
Hence, we take a gauge symmetry broken at an intermediate scale
as a reference model in the following.

\subsection{Evolution of left-right asymmetry}

The evolutions of the AD-fields $\phi$ and $\tilde{\nu}_R$ are governed by
the following equations,
\begin{eqnarray}
	\label{eq:evolution_equation_phi}
	\ddot{\phi} + 3H\dot{\phi} + \frac{\partial V}{\partial \phi^*} &=& 0, \\
	\label{eq:evolution_equation_nu}
	\ddot{\tilde{\nu}}_R + 3H\dot{\tilde{\nu}}_R + \frac{\partial V}{\partial \tilde{\nu}_R^*}
	&=& 0.
\end{eqnarray}
After the inflation ends,
the energy density of the universe is dominated by the coherent oscillation of the inflaton.
Hence, the Hubble parameter varies in time as $H = 2 /(3t)$.
The AD-fields start coherent oscillation before the reheating ends,
since the negative Hubble-induced soft masses are decreased
and effective masses for them dominate the scalar potential.
As a result, the left-right asymmetry is
generated through the evolution of the AD-fields during $ H_R < H < H_{\rm inf}$
where $H_R$ is the Hubble parameter when the reheating ends.

Left-handed lepton number $L^{(L)}$ and right-handed lepton number $L^{(R)}$
are given by
\begin{eqnarray}
	\label{eq:LH-lepton-number}
	L^{(L)}
	&=& \frac{i}{2}(\dot{\phi^*}\phi - \phi^*\dot{\phi}), \\
	\label{eq:RH-lepton-number}
	L^{(R)}
	&=& -i(\dot{\tilde{\nu}}_R^*\tilde{\nu}_R - \tilde{\nu}_R^*\dot{\tilde{\nu}}_R),
\end{eqnarray}
where note that the right-handed neutrino is strictly anti-neutrino.
From Eqs.~(\ref{eq:evolution_equation_phi}) and (\ref{eq:evolution_equation_nu}),
the evolution equations of the total lepton number $n_L \equiv L^{(L)} + L^{(R)}$ 
and the left-right asymmetry $n_{LR} \equiv L^{(L)} - L^{(R)}$
are given by
\begin{eqnarray}
	\label{eq:evolution_total_lepton_number}
	\frac{d n_L}{dt} + 3Hn_L &=& 0, \\
	\label{eq:evolution_left-right_asymmetry}
	\frac{d n_{LR}}{dt} + 3H n_{LR} 
	&=& 4{\mathrm{Im}} \left[ y_\nu(A m_{3/2} + a H) \phi^2\tilde{\nu}_R \right] .
\end{eqnarray}
Here, of course, it can be seen that the total lepton number is conserved,
while the left-right asymmetry is produced by
CP-violating A-terms of the neutrino Yukawa coupling.

We can understand the evolution of $n_{LR}$
by Eq.~(\ref{eq:evolution_left-right_asymmetry}).
In the early universe at $H > m_{3/2}$,
the phase direction of the potential of $\phi$ and $\tilde{\nu}_R$ is dominated
by the Hubble-induced A-term, and it is expected that the phases of AD-fields 
are almost fixed at one of the minima.
This allows us to drop the Hubble-induced A-term
from Eq.~(\ref{eq:evolution_left-right_asymmetry}).
Therefore, this equation is reduced to
\begin{eqnarray}
	\label{eq:evolution_left-right_asymmetry_t}
	\frac{d}{dt} \left( \frac{n_{LR}}{H^2} \right)
	= \frac{4 {\mathrm{Im}}(y_\nu A m_{3/2} \phi^2\tilde{\nu}_R )}{H^2}.
\end{eqnarray}
Here, if the source term of Eq.~(\ref{eq:evolution_left-right_asymmetry_t})
does not oscillate and scales in terms of $t$ as
${\mathrm{Im}} ( y_\nu A m_{3/2} \phi^2\tilde{\nu}_R ) / H^2 \propto t^{\gamma + 2}$,
the solution is derived as
\begin{eqnarray}
	\label{eq:scaling_solution_left-right_asymmetry}
	\frac{ n_{LR} }{ H^2 } = \frac{8}{3 H^3 (\gamma+3)}
	{\mathrm{Im}}\left( y_\nu A m_{3/2} \phi^2\tilde{\nu}_R \right) + C ,
\end{eqnarray}
where $C$ is given by
\begin{eqnarray}
	\label{eq:integration_constant}
	C = \frac{n_{LR,0}}{H_0^2} - \frac{8}{3H_0^3 (\gamma+3)}
	{\mathrm{Im}}\left( y_\nu A m_{3/2} \phi_0^2 \tilde{\nu}_{R0} \right).
\end{eqnarray}
Here, the subscript $`0$' refers to the time when the source term begins to evolve as 
${\mathrm{Im}}\left( y_\nu A m_{3/2} \phi^2\tilde{\nu}_R \right) / H^2 \propto t^{\gamma + 2}$.

If $ \gamma > -3$,
the first term of Eq.~(\ref{eq:scaling_solution_left-right_asymmetry}) grows,
and the left-right asymmetry normalized by $H^2$ increases as
\begin{eqnarray}
	\label{eq:growing_solution_left-right_asymmetry}
	\frac{n_{LR}}{H^2} = \frac { 8 } { 3H^3 ( \gamma+3 ) }
	{\mathrm{Im}}\left( y_\nu A m_{3/2} \phi^2\tilde{\nu}_R \right).
\end{eqnarray}
where we assume the initial value $C$ is negligible. 
On the other hand, for $\gamma < -3$, 
the first term of Eq.~(\ref{eq:scaling_solution_left-right_asymmetry})
is attenuated as $t$ is increased.
Hence, the asymmetry approaches quickly the constant value, $C$.
As a result, the asymmetry is not increased significantly in this era.

If the source term is oscillating around zero,
it is expected that the asymmetry does not increase significantly,
since positive and negative contributions are almost canceled each other.
However, the left-right asymmetry can grow even after $\phi$ starts oscillation
if $\tilde{\nu}_R$ is fixed and the trajectory of $\phi$ has high ellipticity,
whose major axis is misaligned from the direction determined by the low-energy A-term.
In this case, the average of the source term,
which is proportional to ${\mathrm{Im}}(\phi^2)$, over one oscillation cycle of $\phi$ 
is almost never canceled.
This case is realized for example in the case that
the phase of $\phi$ is determined by the Hubble-induced A-term
even after the oscillation starts.
In other words, the oscillation begins when $H \gg m_{3/2}$.
Even for these case, the evolution of the left-right asymmetry is estimated by
the scaling of the average of the source term over one oscillation cycle.

\subsection{Evolution of $\phi$ and $\tilde{\nu}_R$}

After the inflation ends, the AD-fields $\phi$ and $\tilde{\nu}_R$ almost stay in
the initial values of $|\phi| = M_{I_L}$ and $|\tilde{\nu}_R| = M_{I_R}$
until they are destabilized.
We define $H_{\rm osc}$ as the Hubble parameter when $\phi$ begins oscillation,
\begin{eqnarray}
	H_{\mathrm{osc}} = \max \{ H_i \},
\end{eqnarray}
where $H_i$ is roughly estimated by the Hubble parameter
when a potential term $V_i$ dominates over the negative Hubble-induced soft mass,
$V_i(H_i) > H_i^2 M_{I_L}^2$.
Accordingly, the evolution after the destabilization
depends on which term destabilizes the AD-fields.
Hence, we discuss the evolution of the AD-fields
divided by four destabilization terms.

\begin{itemize}
	\item Low-energy soft SUSY breaking masses:
	$V_{\mathrm{soft}}= m_\phi^2|\phi|^2 + m_{\tilde{\nu}_R}^2 |\tilde{\nu}_R|^2$ 
	
	These terms dominate over the Hubble-induced mass terms 
	at $H_{\mathrm{soft}} \sim m_\phi (m_{\tilde{\nu}_R})$.
	Here we can safely neglect A-term contributions for $\phi$,
	since $y_\nu M_{I_L}<m_{3/2}$ and $y_\nu M_{I_R}<m_{3/2}$,
	otherwise quartic terms dominate the dynamics of AD-fields.
	\item Quartic terms:
	$V_{\mathrm{quar}} = y_\nu^2|\phi|^4/4 + y_\nu^2|\phi|^2|\tilde{\nu}_R|^2$
	
	These terms dominate over the Hubble-induced mass term 
	at $H_{\mathrm{quar}} \sim \max \{y_\nu M_{I_L} , y_\nu M_{I_R} \}$.
	In this destabilization case, the Hubble-induced A-term can give a contribution
	comparable to quartic terms.
	Therefore, this contribution should be taken into account.
	\item Thermal-log terms:
	$ V_{\mathrm{log}} = \sum_{f_k|\phi|>T} a_{\rm th} \alpha_s^2(T)T^4 
	\ln (|\phi|^2 / T^2 )$
	
	These terms appear for $H < H_{{\mathrm{thr}},t} = f_t^4 M_{I_L}^4 / (T_R^2 M_{\mathrm{Pl}})$
	and dominate over the Hubble-induced soft mass of $\phi$ for
	$H < \sum a_{\mathrm{th}} \alpha_s^2 T_R^2M_{\mathrm{Pl}}/M_{I_L}^2$.
	Since the former condition is weak in the parameter region we consider,
	$H_{\mathrm{log}} $ is estimated as
	\begin{eqnarray}
		H_{\mathrm{log}} \sim 
		\sum	a_{\mathrm{th}} \frac{ \alpha_s^2 T_R^2M_{\mathrm{Pl}}}{M_{I_L}^2}.
	\end{eqnarray}
	On the other hand, the destabilization of $\tilde{\nu}_R$
	comes from the low-energy soft SUSY breaking mass
	since the amplitude of the oscillation of $\phi$
	decreases quickly after the oscillation of $\phi$ begins
	and the effective mass of $\tilde{\nu}_R$, $y_\nu^2 |\phi|^2 |\tilde{\nu}_R|^2$,
	also decreases quickly.
	As a result, the destabilization of $\tilde{\nu}_R$
	occurs at $H \sim m_{\tilde{\nu}_R}$.	
	\item Thermal-mass terms: $V_{\rm th} = \sum_{f_k|\phi|<T} c_kf_k^2T^2|\phi|^2$
	
	These terms appear if 
	$H > H_{{\mathrm{thr}},k} = f_k^4M_{I_L}^4/ (T_R^2M_{\mathrm{Pl}}) $
	for degrees $k$.
	This effective mass dominates over the Hubble-induced mass term
	at $H \sim H_{{\mathrm{th}},k} = (c_k^2 f_k^4 T_R^2 M_{\mathrm{Pl}})^{1/3}$.
	Therefore, the Hubble parameter which themal-mass terms destabilize $\phi$
	is given by
	\begin{eqnarray}
		H_{\mathrm{th}} \sim \max \{ H_{{\mathrm{th}},k} \}.
		~~~(H_{{\rm th},k} > H_{{\mathrm{thr}},k}  )
	\end{eqnarray}
	Once thermal-mass terms dominate the potential,
	the amplitude of the oscillation of $\phi$ scales as $|\phi| \propto H^{7/8}$
	and the temperature scales as $T \propto H^{1/4}$.
	Therefore, the condition $f_k|\phi| < T$ is kept for subsequent evolution
	and thermal-mass terms do not become ineffective
	until the temperature becomes very low.
	In this case $\tilde{\nu}_R$ is destabilized by the low-energy soft mass
	at $H \sim m_{\tilde{\nu}_R}$ as the thermal-log case.
\end{itemize}
Note that $M_{I_R}$ does not have any effect on the evolution of the AD-fields
for destabilization by thermal effects.

\begin{figure}[t]
	\begin{center}
		\scalebox{.45}{\includegraphics{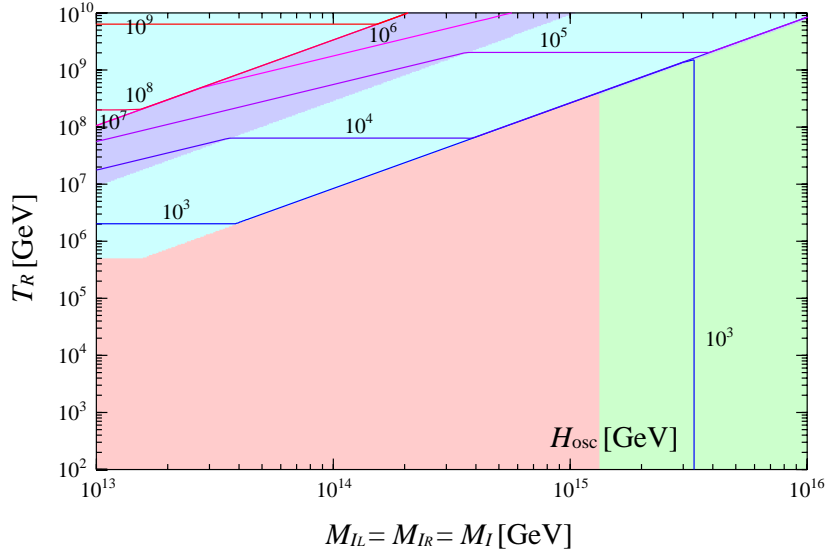}}
		\caption{\small Colored regions indicate which term destabilizes $\phi$:
		low-energy soft masses (red), quartic terms (green),
		thermal-log terms (dark blue), thermal-mass terms (light blue).
		We also show $H_{\mathrm{osc}}$ in the $M_I$-$T_R$ plane,
		where $M_{I_R} = M_{I_L}=M_I$ is assumed.}
		\label{fig:destabilize}
	\end{center}
\end{figure}

In summary of this discussion about the destabilization,
we show which terms destabilize $\phi$ in Fig.~\ref{fig:destabilize};
low-energy soft masses (red), quartic terms (green),
thermal-log terms (dark blue), thermal-mass terms (light blue).
We also show $H_{\mathrm{osc}}$ in the $M_I$-$T_R$ plane,
where $M_{I_R} = M_{I_L}=M_I$ is assumed.
Other parameters are taken as
$y_\nu=1.0\times10^{-12},~m_\phi=m_{\tilde{\nu}_R}=m_{3/2}=1{\mathrm{TeV}},
~c_\phi=c_{\tilde{\nu}_R}=1.0,A=e^{0.6i},~a=-1.0$.
As we will discuss later, since the left-right asymmetry generation
depends on $H_{\mathrm{osc}}$ sensitively,
it is necessary to derive the accurate value of $H_{\mathrm{osc}}$
in order to reduce uncertainty in estimate of the resultant baryon number asymmetry.
According to numerical calculations,
we used following values:
$H_{\mathrm{soft}} = 0.4\times m_\phi,~$
$H_{\mathrm{quar}} = 0.3\times{\mathrm{max}}\{{y_\nu M_{I_L}, y_\nu M_{I_R}}\},~$
$H_{\mathrm{log}} = 0.2\times \sum a_{\mathrm{th}} \alpha_s^2 
T_R^2 M_{\mathrm{Pl}}/M_{I_L}^2,~$
$H_{\mathrm{th}} = 0.2\times {\mathrm{max}}
\{ H_{{\mathrm{th}},k} \theta (H_{{\mathrm {th}},k} - H_{{\mathrm{thr}},k})\}$.
Note that thermal-mass terms appear if $H > H_{{\mathrm{thr}},k}$.
Hence, those terms can dominate the potential
only if $H_{{\rm th},k} > H_{{\mathrm{thr}},k}$.
By this condition, $H_{\rm osc}$ is changed discretely
on lower boundaries in thermal-mass terms regions.

In the following, we consider the evolutions of the AD-fields
and estimate the left-right asymmetry for these four cases in turn.

\subsubsection*{Low-energy soft SUSY breaking masses}

After the AD-fields are destabilized by low-energy soft breaking mass terms
and the origin becomes the minimum of the potential,
$\phi$ begins coherent oscillation, whose amplitude scales with $H$.
On the other hand,
new minimum of $\tilde{\nu}_R$ appears due to the low-energy A-term contribution,
\begin{eqnarray}
	\label{eq:minimum-nu_soft}
	|\tilde{\nu}_R|_{\mathrm{min}} \simeq \frac{y_\nu |A| |\phi^2|}{m_{3/2}} ,
\end{eqnarray}
where $m_{\tilde{\nu}_R} \sim m_{3/2}$.
If $M_{I_L} < M_c \equiv \sqrt{m_{3/2}M_{I_R}/ (y_\nu|A|) }$,
this minimum is smaller than
$M_{I_R}$ at $H=H_{\mathrm{soft}}$,
therefore $\tilde{\nu}_R$ is also destabilized.
After the destabilization of the AD-fields,
$\tilde{\nu}_R$ begins oscillation around this minimum.
However, since this minimum scales as $|\tilde{\nu}_R|_{\rm min} \propto H^2$
and the amplitude of the oscillation of $\tilde{\nu}_R$ scales with $H$,
the difference between the origin and $|\tilde{\nu}_R|_{\rm min}$ becomes negligible soon.
Consequently, the source term in Eq.~(\ref{eq:evolution_left-right_asymmetry_t})
begins oscillation at $H\sim m_{3/2}$ and
the net effect of the source term decreases to negligibly small with a cancelation.
Hence, the left-right asymmetry is fixed once the AD-fields begin oscillation.
If we take the difference between the origin and $|\tilde{\nu}_R|_{\rm min}$ into account,
the scaling of the source term in Eq.~(\ref{eq:evolution_left-right_asymmetry_t})
is estimated as $\gamma=-4$
since $|\tilde{\nu}_R|$ averaged over one oscillation cycle,
which is estimated by the position of the minimum, scales with $|\phi^2|$.
As a result, this effect attenuates soon as discussed above.

If  $M_{I_L} > M_c$,
the minimum is placed at $|\tilde{\nu}_R|_{\mathrm{min}} > M_{I_R}$
when $\phi$ is destabilized.
This means that $\tilde{\nu}_R$ stays at initial value as $|\tilde{\nu}_R| = M_{I_R} $
even after the destabilization of $\phi$ occurs.
Then, when the amplitude of the oscillation of $\phi$ becomes $|\phi| = M_c$,
i.e., $|\tilde{\nu}_R|_{\mathrm{min}} < M_{I_R}$,
$\tilde{\nu}_R$ begins oscillation.
After $\tilde{\nu}_R$ is destabilized,
it oscillates around the minimum as discussed in the previous paragraph.
Note that the oscillation of $\tilde{\nu}_R$ begins at $H < m_{3/2}$ in this case.
However, the left-right asymmetry does not change significantly during $H < m_{3/2}$.
This is because there are no driving forces for phase direction,
since the low-energy A-term dominates the phase direction of the potential.
Therefore, we can estimate the left-right asymmetry,
by assuming that it is fixed at $H\sim m_{3/2}$.

As discussed above, the left-right asymmetry
is fixed after $\phi$ begins oscillation.
Hence, the left-right asymmetry-to-entropy ratio after the reheating
is estimated from the Hubble parameter $H_{\mathrm{soft}}$ 
\begin{eqnarray}
	\label{eq:asymmetry_soft-mass_destabilize}
	\frac{n_{LR}}{s} \simeq \frac{2}{9}
	\frac{y_\nu|A| M_{I_L}^2M_{I_R} m_{3/2}T_R}{H_{\mathrm{soft}}^3 M_{\mathrm{Pl}}^2}\delta_{\rm eff}
\end{eqnarray}
where $\delta_{\rm eff} \lesssim 1 $ is the phase factor of the source term
in Eq.~(\ref{eq:evolution_left-right_asymmetry_t}).
Since Eq.~(\ref{eq:asymmetry_soft-mass_destabilize}) depends on $H_{\mathrm{soft}}^{-3}$,
it is necessary to estimate $H_{\mathrm{soft}}$ accurately in order to reduce
uncertainty in the estimate of $n_{LR}/s$.

\begin{figure}[p]
	\begin{center}
		\scalebox{.7}{ \includegraphics{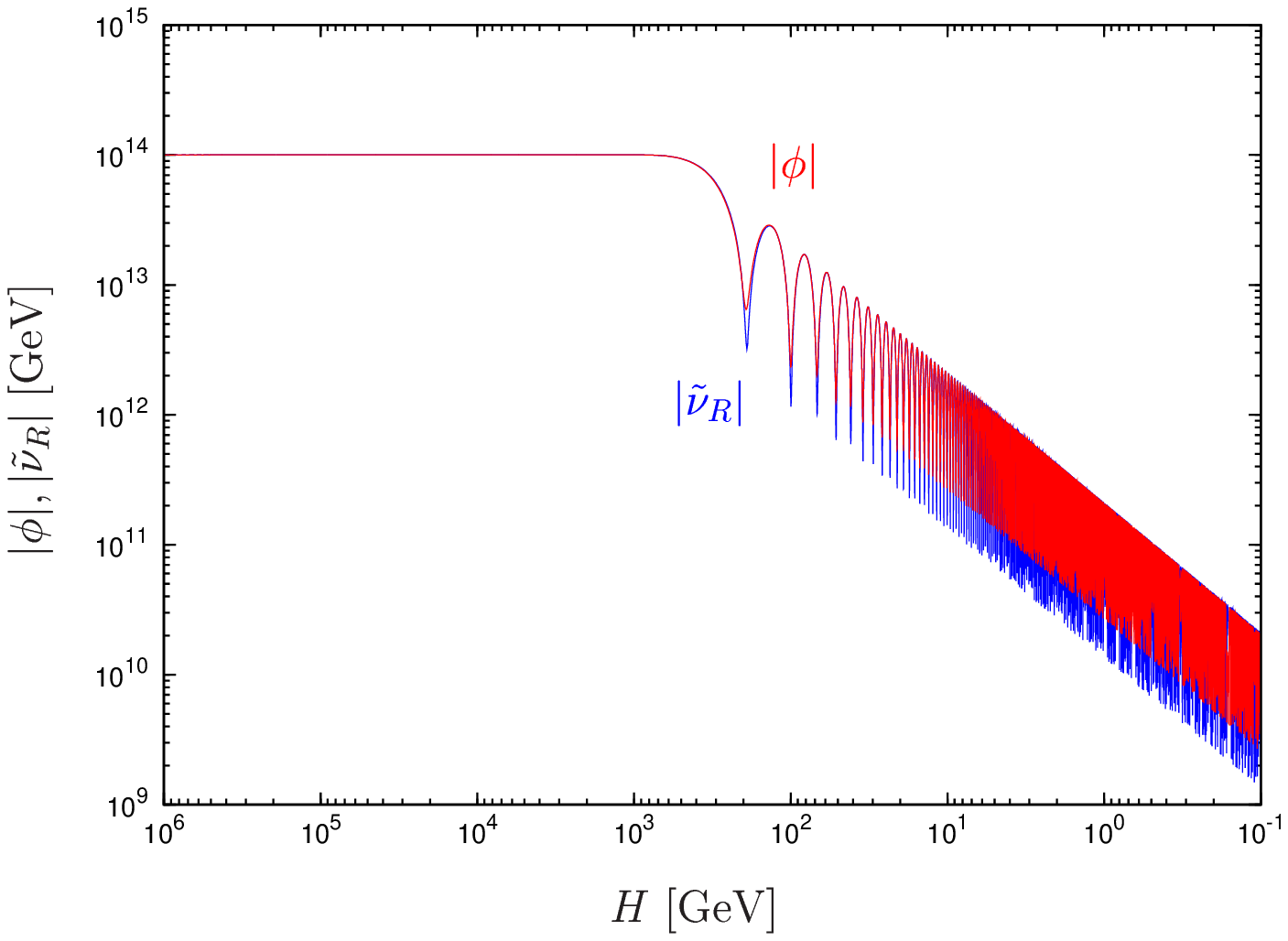}} \\
		\vspace{5mm}
		\scalebox{.7}{ \includegraphics{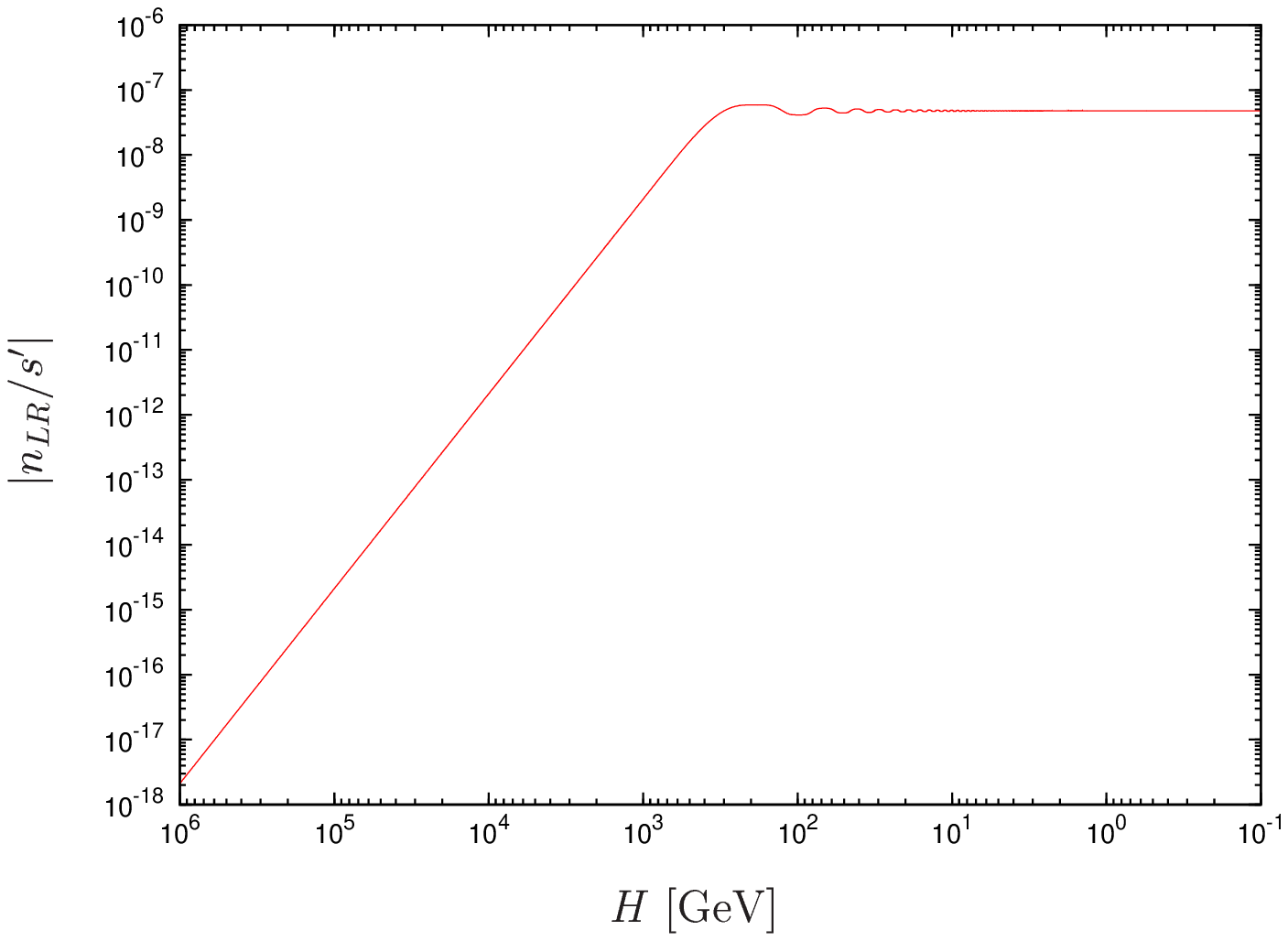}}
		\caption{\small Evolutions of the AD-fields in their magnitude (upper) and
		the left-right asymmetry (lower) are shown in the case that
		low energy soft SUSY breaking mass terms destabilize the AD-fields.
		Parameters are taken as	$m_\phi = m_{\tilde{\nu}_R} = m_{3/2} = 1{\mathrm{TeV}},
		~y_\nu= 10^{-12}, c_\phi = c_{\tilde{\nu}_R}= 1.0, ~A=e^{0.6i},~a = -1.0$,
		$T_R=10^5{\mathrm{GeV}}$, and 
		$ M_{I_L}= M_{I_R} = M_I =10^{14}{\mathrm{GeV}}$.
		An entropy parameter $s'$ is defined as $s' \equiv 4H^2 M_{\mathrm{Pl}}^2 / T_R$.}
		\label{fig:evolution-fields-soft}
	\end{center}
\end{figure}

The evolutions of the magnitude of the AD-fields and the asymmetry are shown 
in Fig.~\ref{fig:evolution-fields-soft}.
In these figures, we take the following parameter values
$m_\phi = m_{\tilde{\nu}_R} = m_{3/2} = 1{\mathrm{TeV}},
~y_\nu= 10^{-12}, c_\phi = c_{\tilde{\nu}_R}= 1.0, ~A=e^{0.6i},~a = -1.0,
~T_R=10^5{\mathrm{GeV}}$, and 
$ M_{I_L}= M_{I_R} = M_I =10^{14}{\mathrm{GeV}}$.
We define an entropy parameter as $s' \equiv 4H^2M_{\mathrm{Pl}}^2 / T_R$,
which is normalized
in order to give the left-right asymmetry-to-entropy ratio $n_{LR}/s$ after the reheating ends.
From these figures, we can confirm that
the AD-fields begin oscillation at $ H \sim m_\phi \sim m_{\tilde{\nu}_R}$
and the left-right asymmetry is fixed after the AD-fields begin oscillation.

\subsubsection*{Quartic terms}

In this case, the evolutions of the AD-fields are very complicated,
because we can not neglect A-term contributions.
With the Hubble-induced soft terms and quartic terms,
$\phi$ and $\tilde{\nu}_R$ are destabilized at $H_1$ and $H_2$, respectively,
\begin{eqnarray}
	\label{eq:quartic-destabilize}
	H_1 &\simeq& \frac{1}{c_\phi}\left( -|a|y_\nu M_{I_R}+  y_\nu\sqrt{|a|^2 M_{I_R}^2 
	+ \left(M_{I_R}^2 + M_{I_L}^2/2\right) c_\phi} \right), \\
	H_2 &\simeq& \frac{1}{2c_{\tilde{\nu}_R}}\left( -|a|y_\nu M_{I_L}^2 / M_{I_R}
	+  y_\nu M_{I_L} \sqrt{|a|^2 M_{I_L}^2 / M_{I_R}^2 + 4 c_{\tilde{\nu}_R}} \right).
\end{eqnarray}
Both quartic terms contribute to the effective mass of $\phi$,
and $\phi$ is destabilized when one of these terms dominates the potential.
On the other hand, only $y_\nu^2 |\phi|^2|\tilde{\nu}_R|^2$ term contributes to 
the effective mass of $\tilde{\nu}_R$.
Hence, $\phi$ can be destabilized
before the destabilization of $\tilde{\nu}_R$ for $M_{I_L} \ll M_{I_R}$.

For $M_{I_L} \gtrsim M_{I_R}$ after the destabilization,
$\tilde{\nu}_R$ oscillates around the minimum determined by the balance between 
the Hubble-induced A-term and the quartic term,
\begin{eqnarray}
	\label{eq:minimum-nu_quartic}
	|\tilde{\nu}_R|_{\mathrm{min}}=\frac{|a|H}{y_\nu}. ~~~ (H>m_{3/2})
\end{eqnarray}
Hence, for $\tilde{\nu}_R$ destabilization,
two conditions are required.
The one is that the effective mass from the quartic term
dominates over the Hubble-induced negative mass and
the other is that $|\tilde{\nu}_R|_{\min} < M_{I_R}$.
After the destabilization, $\tilde{\nu}_R$ begins oscillation
around the instantaneous minimum, Eq.~(\ref{eq:minimum-nu_quartic}).

In addition, if $y_\nu M_{I_R}/|A| > m_{3/2}$,
the destabilization of $\tilde{\nu}_R$ does not occur for $H > m_{3/2}$ and
$\tilde{\nu}_R$ remains stabilized at the initial value
by the low-energy soft SUSY breaking A-term after $H \sim m_{3/2}$.
While the quartic term still dominates over the low-energy soft mass for $H < m_{3/2}$,
the minimum is fixed at
\begin{eqnarray}
	\label{eq:minimum-nu_quartic-soft}
	|\tilde{\nu}_R|_{\mathrm{min}}=\frac{|A|m_{3/2}}{y_\nu}. ~~~ (H < m_{3/2})
\end{eqnarray}
After the amplitude of oscillation of $\phi$ decreases to $|\phi| = M_c$,
the low-energy soft mass dominates the potential.
Then, as discussed in the previous case, the minimum is given by 
\begin{eqnarray}
	\label{eq:minimum-nu_soft2}
	|\tilde{\nu}_R|_{\mathrm{min}}=\frac{y_\nu |A| |\phi|^2}{m_{3/2}},
\end{eqnarray}
and the subsequent evolution is also the same as in the previous case.

If $M_{I_L} \ll M_{I_R}$,
$\phi$ is destabilized at $H\sim y_\nu M_{I_R}/\sqrt{2}$
before the destabilization of $\tilde{\nu}_R$.
Since $\tilde{\nu}_R$ takes constant value, the effective mass of $\phi$ is also constant.
Hence, the amplitude of the oscillation of $\phi$ scales with $H$.
In this case, since both quartic terms and SUSY breaking A-terms decrease quickly,
$| \tilde{\nu}_R |$ stays at $M_{I_R}$
until the low-energy soft mass dominates over the Hubble-induced mass\footnote{
After $\tilde{\nu}_R$ is destabilized, $\phi$ may begin chaotic evolution,
because the effective mass of $\phi$ from the quartic term $y_\nu^2|\phi|^2|\tilde{\nu}_R|^2$
is oscillating with very large amplitude and this results in instability of $\phi$.
However, since this takes place well after the left-right asymmetry is fixed,
this evolution does not have any effect on the estimate of the asymmetry.
}.

In both cases, $M_{I_L} \gtrsim M_{I_R}$ and $M_{I_L} \ll M_{I_R}$,
The left-right asymmetry grows significantly until $H \sim m_{3/2}$.
First, in both cases it continues to increase 
after $\phi$ begins oscillation by quartic terms as mentioned above.
Moreover, the left-right asymmetry does not cease to grow
even during $\tilde{\nu}_R$ oscillation around the minimum (\ref{eq:minimum-nu_quartic}),
because the center of the oscillation of $\tilde{\nu}_R$ is significantly separated from the origin.
The scaling of the source term in Eq.~(\ref{eq:evolution_left-right_asymmetry_t})
is estimated by the amplitude of the oscillation of $\phi$,
which scales with $H^{2/3}$ for the oscillation by quartic terms,
and the minimum of $\tilde{\nu}_R$, $|\tilde{\nu}_R|_{\min} \propto H$.
Therefore, the net left-right asymmetry grows with $\gamma=-7/3$ in this era.
For $H < m_{3/2}$, the low-energy A-term dominates over the Hubble-induced one,
and the trajectory of $\phi$ tends to be aligned to minimize the low-energy A-term.
As a result, the left-right asymmetry does not change significantly for $H < m_{3/2}$.
After the oscillation of $\phi$ is dominated by the low-energy soft mass,
the asymmetry is fixed at a constant value as discussed in the previous case.

However, there is an uncertainty in the estimate of the left-right asymmetry.
The oscillation of the source term
may leave significant effects
if the cancelation is not sufficient.
Even if both $\phi$ and $\tilde{\nu}_R$ are oscillating by quartic terms,
the magnitude of the source term scales as $\gamma = -2$.
This introduces oscillatory evolution of the left-right asymmetry,
though its order of magnitude is not expected to change significantly.
This oscillation of the asymmetry remains
as uncertainty in our estimate.
Note that, if we assume $\phi$ and $\tilde{\nu}_R$ are at most GUT scale, 
$M_{\mathrm{GUT}} \sim 10^{16}{\mathrm{GeV}}$,
this period is rather short.
The AD-fields are destabilized at $H = H_{\mathrm{quar}} \lesssim 10^4{\mathrm{GeV}}$
for $y_\nu \sim 10^{-12}$,
and the low-energy soft masses govern the evolution of AD-fields
soon after $H \sim m_{3/2}$.
Although the following estimate of the left-right asymmetry is rough,
deviations from numerical results
are within an order of magnitude in the parameter region where we focus on.
In order to derive an accurate result,
numerical calculation is required.

The left-right asymmetry-to-entropy ratio after the reheating ends is estimated as
\begin{eqnarray}
	\label{eq:asymmetry_quartic_destabilize1}
	\frac{n_{LR}}{s} \simeq
	\frac{M_{I_L}^2 m_{3/2}^\frac{1}{3} T_R }
	{H_{\mathrm{quar}}^\frac{4}{3}M_{\mathrm{Pl}}^2 } |a|^\frac{5}{3} |A|^\frac{1}{3} \delta_{\rm eff}
\end{eqnarray}
for $y_\nu M_{I_R}/ |A| > m_{3/2}$ and $M_{I_L} \gtrsim M_{I_R}$,
\begin{eqnarray}
	\label{eq:asymmetry_quartic_destabilize2}
	\frac{n_{LR}}{s} \simeq
	\frac{2}{5} \frac{y_\nu M_{I_L}^2 M_{I_R} T_R}
	{m_{3/2}^\frac{2}{3} H_{\mathrm{quar}}^\frac{4}{3}
	M_{\mathrm{Pl}}^2} |a|^\frac{5}{3} |A|^{-\frac{2}{3}} \delta_{\rm eff}
\end{eqnarray}
for $y_\nu M_{I_R}/ |A| < m_{3/2}$ and $M_{I_L} \gtrsim M_{I_R}$, and 
\begin{eqnarray}
	\label{eq:asymmetry_quartic_destabilize_R>L}
	\frac{n_{LR}}{s} \simeq \frac{2}{3}
	\frac{y_\nu M_{I_L}^2 M_{I_R} m_{3/2} T_R}
	{H_{\mathrm{quar}}^2 H_{\mathrm{soft}} M_{\mathrm{Pl}}^2} |A| \delta_{\rm eff}
\end{eqnarray}
for $M_{I_L} \ll M_{I_R}$.

\begin{figure}[p]
	\begin{center}
		\scalebox{.7}{\includegraphics{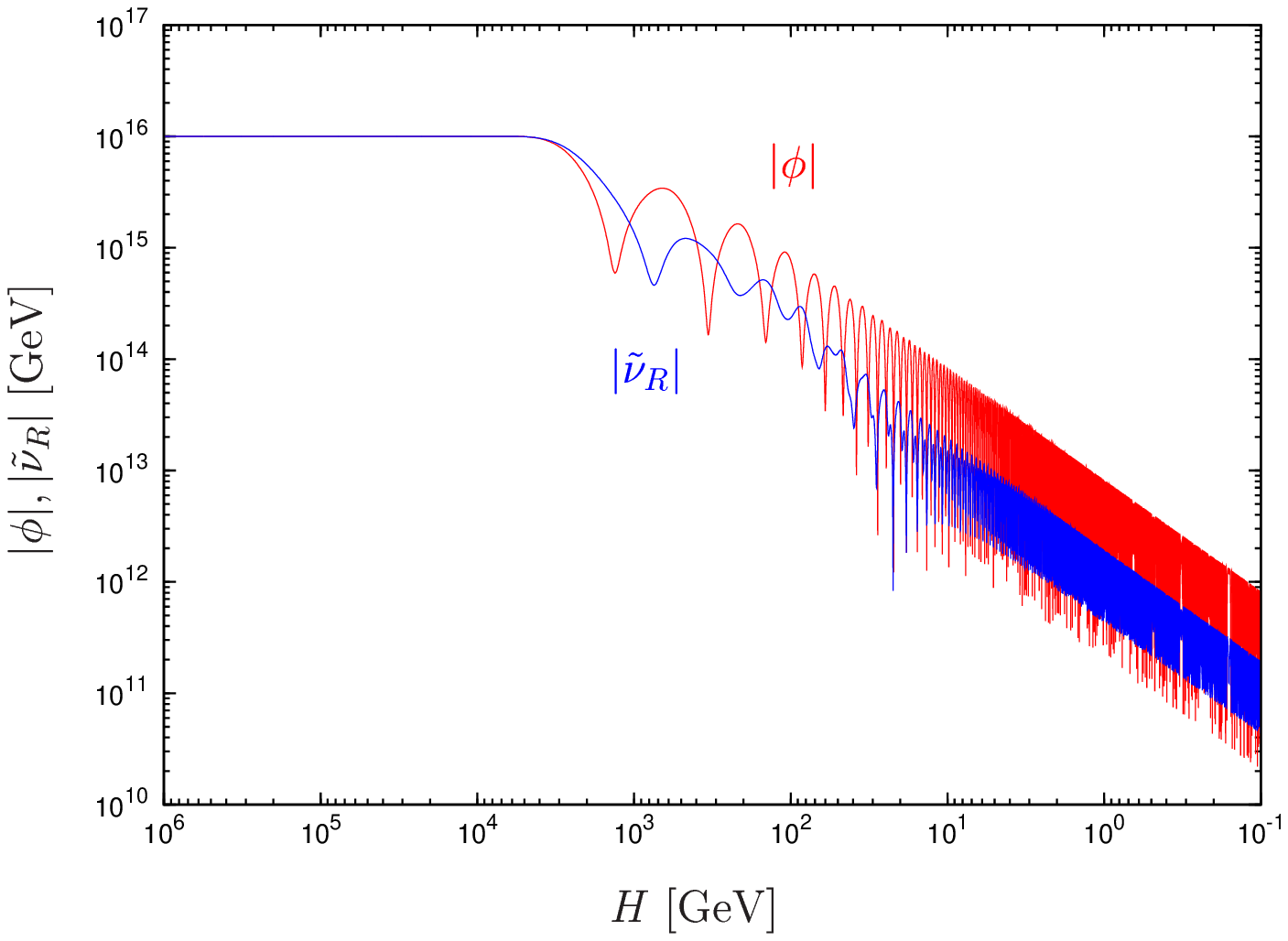}} \\
		\vspace{5mm}
		\scalebox{.7}{\includegraphics{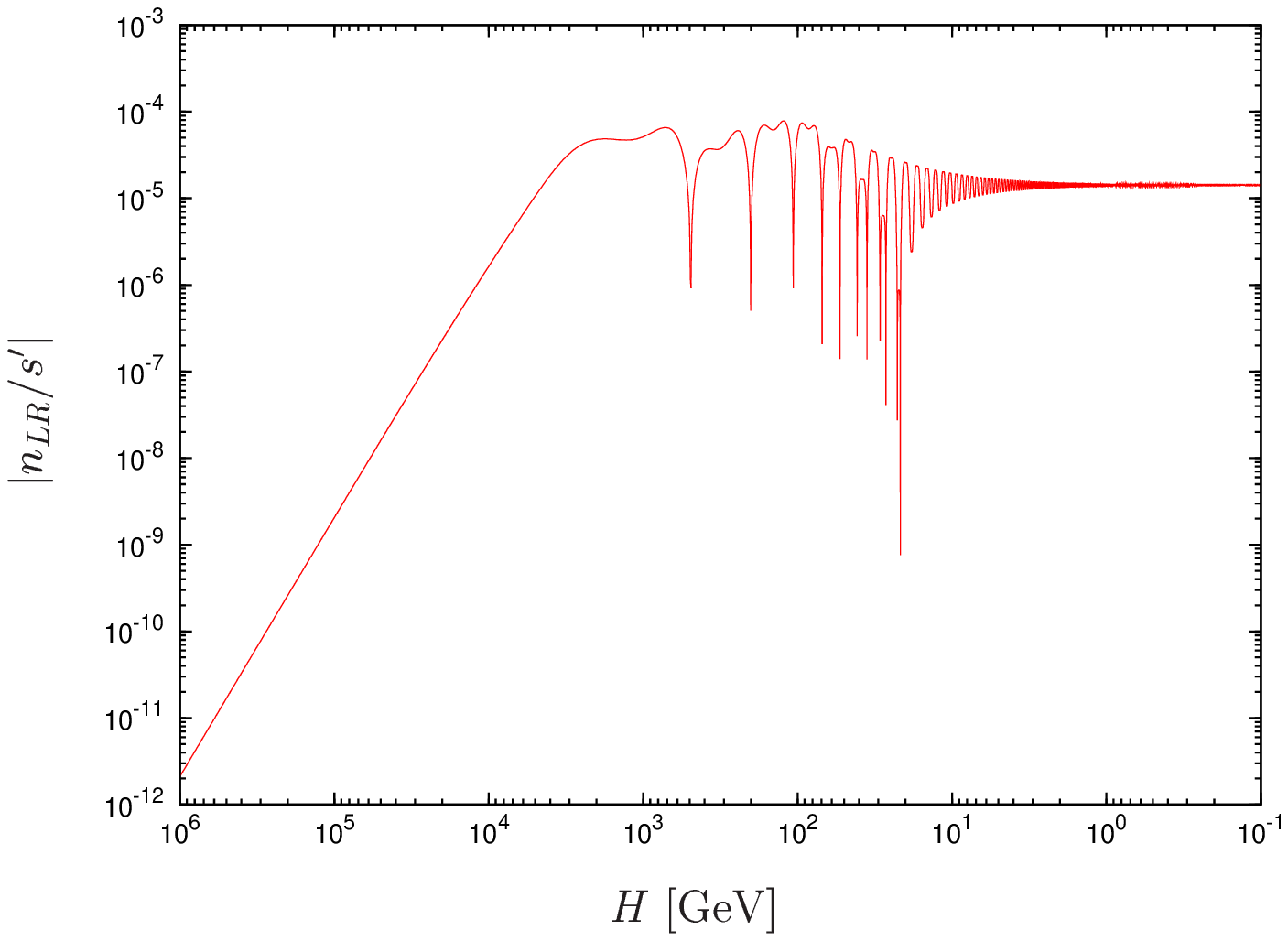}}		
		\caption{\small Evolutions of the AD-fields in their magnitude (upper) and
		the left-right asymmetry (lower) are shown in the case that
		quartic terms destabilize $\phi$.
		Parameters are the same as for in Fig.~\ref{fig:evolution-fields-soft}
		except for $ M_{I_L}= M_{I_R} = M_I =10^{16}{\mathrm{GeV}}$. }
		\label{fig:evolution-fields-quart}
	\end{center}
\end{figure}

The evolutions of the magnitude of the AD-fields and the asymmetry are shown 
in Fig.~\ref{fig:evolution-fields-quart}.
In these figures, parameters are the same as for in Fig.~\ref{fig:evolution-fields-soft}
except for $ M_{I_L}= M_{I_R} = M_I =10^{16}{\mathrm{GeV}}$.
From these figures,
we can see that $\tilde{\nu}_R$ oscillates around the instantaneous minimum
determined by the balance between A-terms and quartic terms.
Then, when $H$ is less than $10 {\mathrm{GeV}}$,
the low-energy soft masses drive the oscillations of the AD-fields.
On the other hand,
we can see that the left-right asymmetry oscillates after the destabilization
and this oscillatory evolution ends after the low-energy soft masses dominates the potential.
We can confirm from this figure
that the resultant asymmetry does not change significantly
in spite of this oscillatory evolution of the asymmetry.

\subsubsection*{Thermal-log terms}

In this case, $\phi$ begins oscillation at $H = H_{\mathrm{log}}$.
Without numerical analysis,
we can hardly trace the evolutions of $\phi$ and ${\tilde{\nu} _R}$
after thermal terms dominate the potential,
since the evolution of $\phi$ is very complicated.
This is because thermal-mass and thermal-log terms 
have comparable size in the epoch when MSSM particles enter the thermal bath in turn.

We summarize typical evolutions of the AD-fields in the following.
First, we assume that all MSSM particles interacting with $\phi$
are not in the thermal bath, $f_k|\phi| > T$,
and thermal-log terms dominate the potential of $\phi$ at $H=H_{\mathrm{log}}$.
By these terms, $\phi$ oscillates and its amplitude is damped fast.
Hence, up quarks and squarks enter the thermal bath soon when $f_u |\phi| \lesssim T$.
Since thermal-mass terms from up quarks and squarks are larger than thermal-log terms,
$c_u f_u^2 T^2 > 0.47\alpha_s^2T^4/|\phi|^2$ by ${\cal O} (\alpha_s^{-2})$,
thermal-mass terms rule the oscillation of $\phi$.
Since the amplitude of the oscillation by thermal-mass terms scales as $H^{7/8}$
and $T \propto H^{1/4}$,
thermal-mass terms redshift faster than thermal-log terms.
Therefore, after a while, thermal-log terms dominate over thermal-mass terms again.
Then, charm quarks and squarks enter the thermal bath next.
In this way, thermal-log terms and thermal-mass terms dominate the potential by turns
until SU(2)$_L$ gauge bosons enter the thermal bath.
Once the SU(2)$_L$ gauge coupling contributes to thermal-mass terms,
thermal-log terms vanish before they dominate over thermal-mass terms again,
since top quarks and squarks enter the thermal bath soon.

\begin{figure}[t]
	\begin{center}
		\scalebox{.8}{\includegraphics{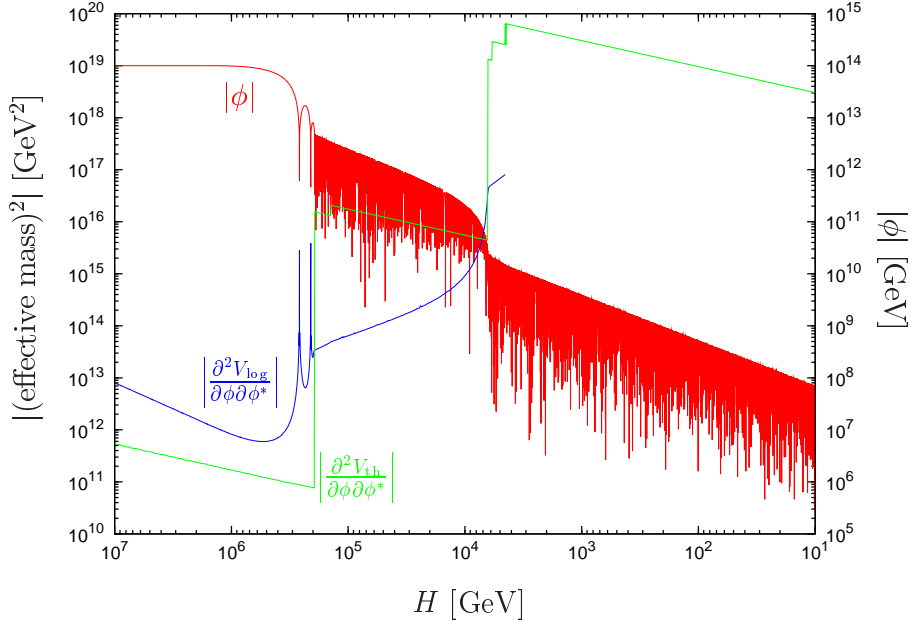}}
		\caption{\small The evolution of $\phi$ 
		after thermal-log terms destabilize $\phi$ at $H\sim10^6{\mathrm{GeV}}$.
		Parameters are taken to be the same as for Fig.~\ref{fig:evolution-fields-soft},
		except for $T_R=10^9{\mathrm{GeV}}$.
		Effective masses of thermal-mass terms 
		$\partial V_{\mathrm{th}}/\partial\phi\partial\phi^*$ and thermal-log terms
		$\partial V_{\mathrm{log}}/\partial\phi\partial\phi^*$,
		are also shown.
		Effects of frequent oscillation of $\phi$
		on these effective masses are erased
		in order to show clearly which is dominant.}
		\label{fig:thermal-evolutions}
	\end{center}
\end{figure}

We show the evolution of $|\phi|$ in Fig.~\ref{fig:thermal-evolutions}.
Parameters are taken to be the same as for Fig.~\ref{fig:evolution-fields-soft},
except for $T_R=10^9{\mathrm{GeV}}$.
Effective masses of thermal-mass terms 
$\partial^2 V_{\mathrm{th}}/\partial\phi\partial\phi^*$ and thermal-log terms
$\partial^2 V_{\mathrm{log}}/\partial\phi\partial\phi^*$
are also shown in this figure.
Here, effects of frequent oscillation of $\phi$ on these effective masses are erased
in order to show clearly which is dominant.
In this case, up (s)quarks enter the thermal bath at $H\sim10^7{\mathrm{GeV}}$.
Hence, they are already in the thermal bath at first.
Thermal-log terms vanish around $H = $ several TeV,
just after all particles relevant to the strong interaction enter the thermal bath.
We can confirm that themal-mass terms and thermal-log terms
dominate the potential by turns from this figure.

It is difficult to trace quantitatively the evolutions of AD-fields in this era,
because of the complicated nature of thermal effects.
Indeed, the amplitude of the oscillation of $\phi$ falls abruptly soon after thermal-log terms
become dominant, rather than scales simply with $H^{3/2}$.
In spite of this complicated evolution,
we can approximate the left-right asymmetry-to-entropy ratio after the reheating ends
by the value at $H = H_{\rm log}$.
One of the reason is that the source term of the left-right asymmetry decreases significantly
because of the abrupt fall of the amplitude of $\phi$ oscillation.
Therefore, the left-right asymmetry can hardly grow from the value at  $H = H_{\rm log}$,
although the source term averaged over oscillations can grow with $\gamma = -7/4$
during $\phi$ is oscillating by thermal-mass terms.
In addition, in most of the parameter region we consider,
the left-right asymmetry does not grow significantly
after top quarks and squarks enter the thermal bath,
because the low-energy soft SUSY breaking mass destabilizes $\tilde{\nu}_R$ soon.

The left-right asymmetry-to-entropy ratio after the reheating ends
is estimated as
\begin{eqnarray}
	\label{eq:asymmetry_therlog_destabilize}
	\frac{n_{LR}}{s} \simeq \frac{2}{9}
	\frac{y_\nu|A| M_{I_L}^2M_{I_R} m_{3/2} T_R}{H_{\mathrm{log}}^3 
	M_{\mathrm{Pl}}^2}\delta_{\rm eff}
	\times Q_{\mathrm{th}}
\end{eqnarray}
where the difference between the value at $H = H_{\rm log}$ and after the reheating ends
is parameterized by $Q_{\rm th}$.
The value of $Q_{\rm th} \gtrsim 1$ is derived by numerical culculations.

\begin{figure}[p]
	\begin{center}
		\scalebox{.7}{\includegraphics{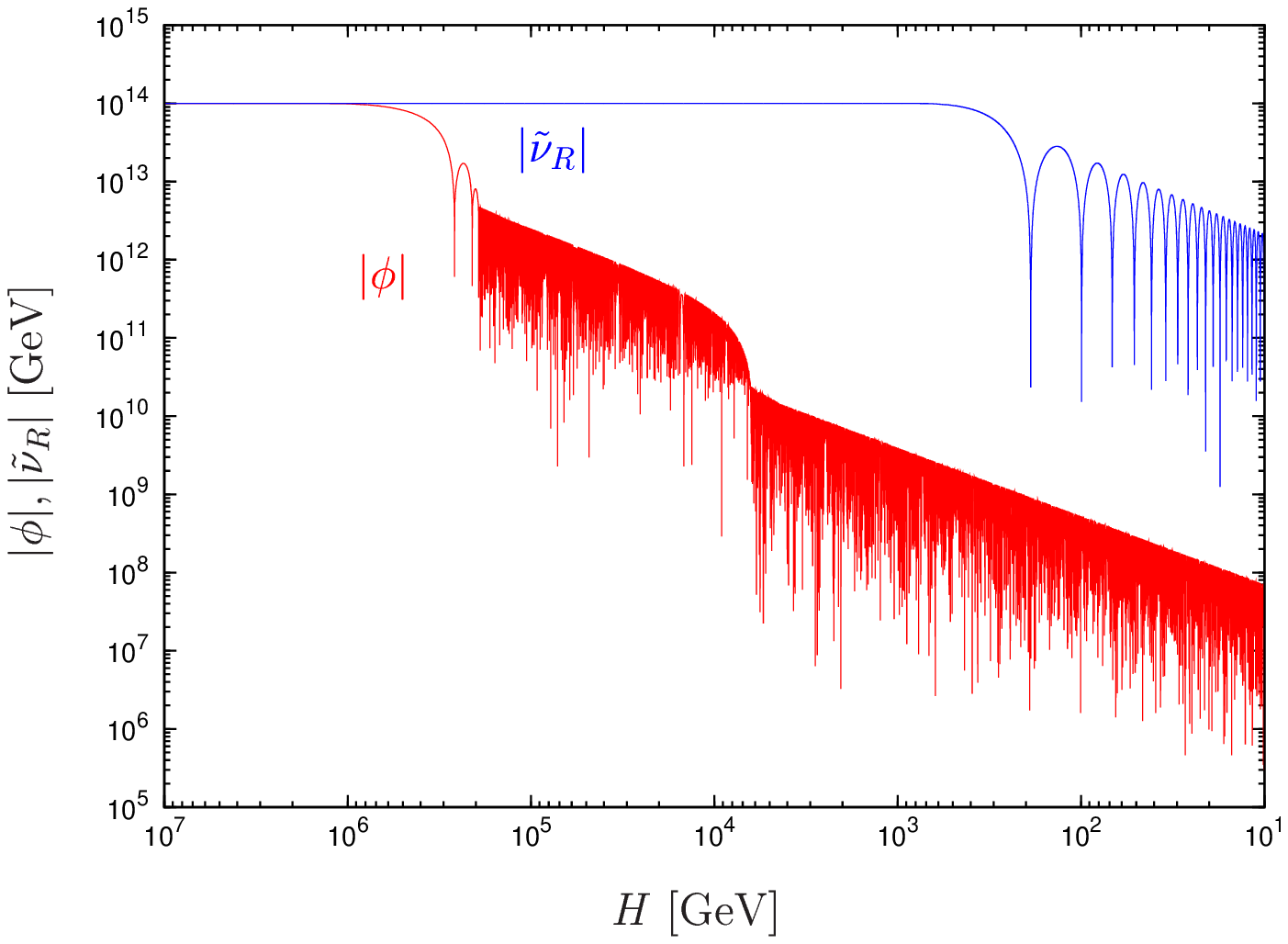}} \\
		\vspace{5mm}
		\scalebox{.7}{\includegraphics{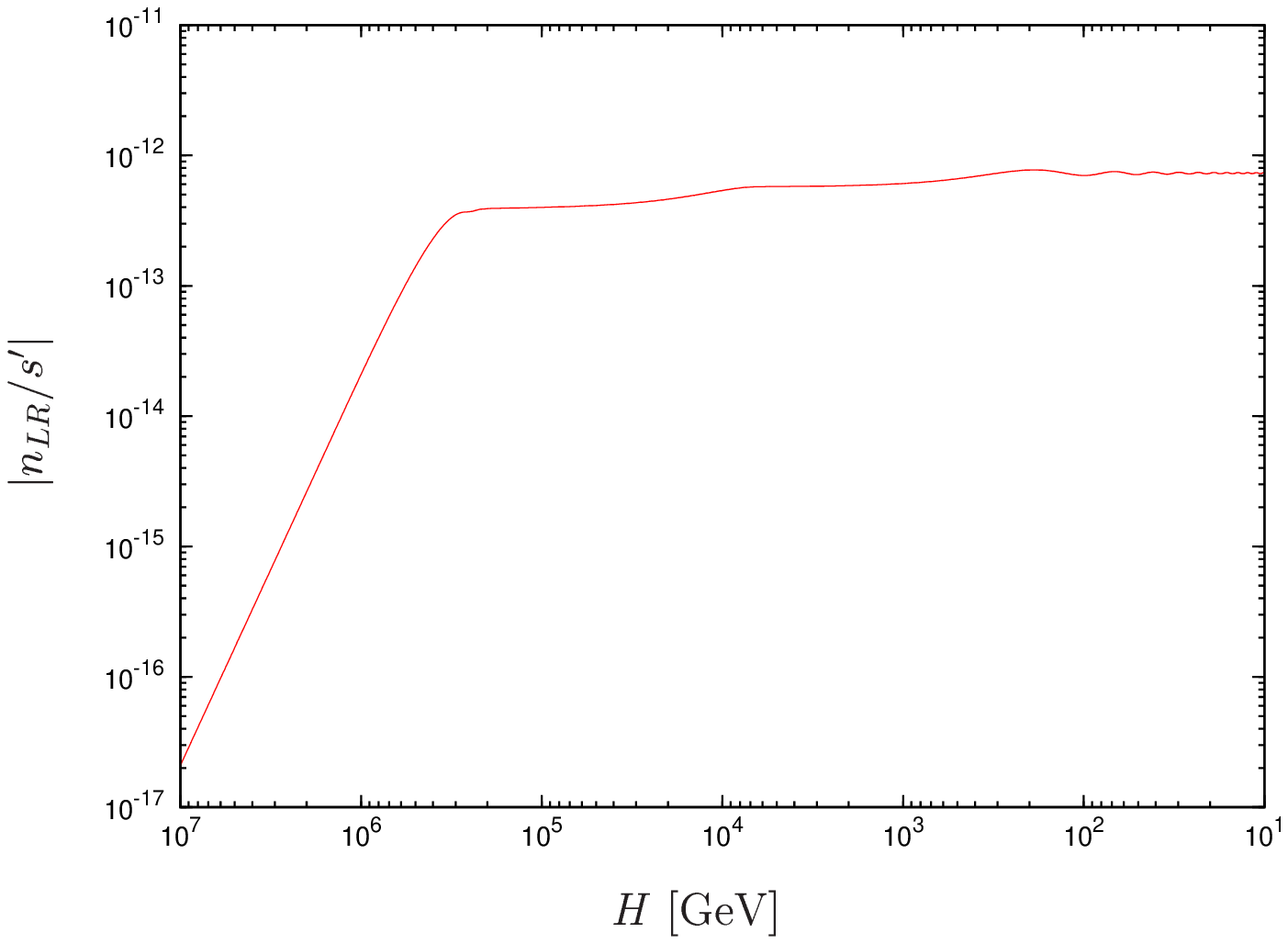}}
		\caption{\small  Evolutions of the AD-fields in their magnitude (upper) and
		the left-right asymmetry (lower) are shown in the case that
		thermal-log terms destabilize $\phi$. 
		Parameters are the same as for Fig.~\ref{fig:thermal-evolutions}.
		}
		\label{fig:evolution-fields-therlog}
	\end{center}
\end{figure}

Figure \ref{fig:evolution-fields-therlog} shows
the evolutions of the magnitudes of the AD-fields and the left-right asymmetry
in the case that thermal-log terms destabilize $\phi$.
parameters are the same as for Fig.~\ref{fig:thermal-evolutions}.
The oscillation of $\phi$ begins at $H = H_{\rm log} \sim 10^6{\mathrm{GeV}}$,
while $\tilde{\nu}_R$ almost remains at the initial value until $H \sim m_{\tilde{\nu}_R}$.
From the lower figure,
We can confirm that the increase of the left-right asymmetry is small
after $\phi$ begins oscillation 
and it is completely fixed after $\tilde{\nu}_R$ begins oscillation.
We can see $Q_{\rm th} \simeq 2$ in this figure.

If thermal effects destabilize the potential early enough,
thermal-mass terms dominate the dynamics for sufficiently long period
after top quarks and squarks participate into the thermal bath.
In this case, the left-right asymmetry can grow significantly during this period.
For example, according to a numerical calculation,
we get $Q_{\mathrm{th}} \sim 10$ if we take $M_I=5\times10^{13}{\mathrm{GeV}}$
and $T_R=10^9{\mathrm{GeV}}$.
However, since destabilization takes place sufficiently early,
the amount of the left-right asymmetry is extremely small:
the result of numerical calculation gives $n_{LR}/s \sim 10^{-14}$ in this example.
Therefore, we can safely ignore these cases.

\subsubsection*{Thermal-mass terms}

In this case, $\phi$ begins oscillation at $H=H_{\mathrm{th}}$, 
while $\tilde{\nu}_R$ almost remains at the initial value until $H\sim m_{\tilde{\nu}_R}$.
In similar to the previous case, 
we can approximate the left-right asymmetry-to-entropy ratio after the reheating ends
by the value at $H = H_{\rm th}$.
The left-right asymmetry-to-entropy ratio is estimated as
\begin{eqnarray}
	\label{eq:asymmetry_thermass_destabilize}
	\frac{n_{LR}}{s} \simeq \frac{2}{9}
	\frac{y_\nu|A| M_{I_L}^2M_{I_R} m_{3/2}T_R}{H_{\mathrm{th}}^3 M_{\mathrm{Pl}}^2}\delta_{\rm eff}
	\times Q_{\mathrm{th}}.
\end{eqnarray}

In some cases, destabilization takes place very late, say, $H\lesssim 10^5{\mathrm{GeV}}$.
Because the low-energy soft SUSY breaking A-term has non-negligible effect,
the evolution of the phase direction can not be ignored.
If this effect is significantly large, 
it results in slow oscillation of the source term around zero,
and therefore the value of the left-right asymmetry 
may be smaller than Eq.~(\ref{eq:asymmetry_thermass_destabilize}).
Since this evolution is very complicated,
we will not discuss the detail.
Instead, we give approximate estimate of the left-right asymmetry 
by Eq.~(\ref{eq:asymmetry_thermass_destabilize}) with $Q_{\mathrm{th}} \lesssim 1$.

On the other hand, if destabilization takes place early enough,
the left-right asymmetry grows drastically.
If we take $M_I=2\times10^{13}{\mathrm{GeV}}$ and $T_R=10^9{\mathrm{GeV}}$,
the result of numerical calculation shows very large growth after $H\sim H_{\rm th}$:
$Q_{\mathrm{th}}\sim10^5$.
However, $n_{LR}/s \sim 10^{-19}$ in this case, because of the early destabilization.
Hence we drop these possibility from our estimate of the asymmetry.

\section{Baryon number asymmetry}

The left-handed lepton number stored in $\phi$
is partially released into the thermal bath after the asymmetry is fixed.
On the other hand, the right-handed lepton number remains in $\tilde{\nu}_R$ condensate
after the electroweak phase transition
due to the smallness of the Yukawa couplings.
For example, it decays through small Yukawa coupling into Higgsino
and left-handed anti-neutrino at $T \lesssim 10{\mathrm{MeV}}$.
Therefore, the left-handed lepton number asymmetry
is transferred to the baryon number asymmetry by the sphaleron process
since it acts only on left-handed leptons.

The chemical equilibrium between leptons and baryons
leads the following ratio \cite{Harvey:1990qw},
\begin{eqnarray}
	\label{eq:baryon_asymmetry}
	B = \frac{8}{23}(B-L^{(L)}) = \frac{8}{23} L^{(R)}.
\end{eqnarray}
Note that $|L^{(L)}| = |L^{(R)} | = |n_{LR}|/2$ due to the total lepton number conservation.
After the electroweak phase transition occurs,
the baryon-to-entropy ratio is conserved since the sphaleron process is ineffective. 
Though $\tilde{\nu}_R$ decays and $L^{(R)}$ is released into other particles,
the baryon number asymmetry is not changed after the electroweak phase transition occurs.

Here, we summarize 
analytic formulae of the baryon-to-entropy ratio generated by this mechanism. 
In the case that the low-energy soft SUSY breaking masses
destabilize the AD-fields,
it is given as
\begin{eqnarray}
	\label{eq:analytic_baryon-to-entropy_soft}
	\frac{n_{B}}{s} &\simeq& 1 \times10^{-8} ~ \left( \frac{y_\nu}{10^{-12}} \right)
	\left(\frac{m_{3/2}}{1{\mathrm{TeV}}}\right)\left(\frac{m_\phi}{1{\mathrm{TeV}}}\right)^{-3}
	\nonumber \\
	&& \times\left(\frac{M_{I_L}}{10^{14}{\mathrm{GeV}}}\right)^2
	\left(\frac{M_{I_R}}{10^{14}{\mathrm{GeV}}}\right)
	\left(\frac{T_R}{10^5{\mathrm{GeV}}}\right)|A|\delta_{\mathrm{eff}}.
\end{eqnarray}
Here we used $H_{\mathrm{soft}} = 0.4\times m_{\phi}$
according to numerical calculations. 
In the case that quartic terms
destabilize the AD-fields,
the asymmetry is estimated as
\begin{eqnarray}
	\label{eq:analytic_baryon-to-entropy_quart_L>R1}
	\frac{n_{B}}{s} &\simeq& 7\times10^{-5} ~ \left( \frac{y_\nu}{10^{-12}} \right)^{-\frac{4}{3}}
	\left(\frac{m_{3/2}}{1{\mathrm{TeV}}}\right)^\frac{1}{3}
	~~~~~~~~~~~~~~~~~~~~~~\nonumber\\
	&& \times \left(\frac{M_{I_L}}{10^{16}{\mathrm{GeV}}}\right)^\frac{2}{3}
	\left(\frac{T_R}{10^5{\mathrm{GeV}}}\right)
	|A|^\frac{1}{3}|a|^{\frac{5}{3}}\delta_{\mathrm{eff}}
\end{eqnarray}
for $M_{I_L} \gtrsim M_{I_R}$ and $y_\nu M_{I_R}/|A| < m_{3/2}$,
\begin{eqnarray}
	\label{eq:analytic_baryon-to-entropy_quart_L>R2}
	\frac{n_{B}}{s} &\simeq& 3 \times 10^{-6} ~ \left( \frac{y_\nu}{10^{-12}} \right)^{-\frac{1}{3}}
	\left(\frac{m_{3/2}}{1{\mathrm{TeV}}}\right)^{-\frac{2}{3}}
	~~~~~~~~~~~~~~~~~~~~~~\nonumber\\
	&& \times\left(\frac{M_{I_L}}{10^{16}{\mathrm{GeV}}}\right)^\frac{2}{3}
	\left(\frac{M_{I_R}}{10^{14}{\mathrm{GeV}}}\right)
	\left(\frac{T_R}{10^5{\mathrm{GeV}}}\right)
	|A|^{-\frac{2}{3}} |a|^{\frac{5}{3}}\delta_{\mathrm{eff}}
\end{eqnarray}
for $M_{I_L} \gtrsim M_{I_R}$ and $y_\nu M_{I_R}/|A| < m_{3/2}$, and 
\begin{eqnarray}
	\label{eq:analytic_baryon-to-entropy_quart_R>L}
	\frac{n_{B}}{s} &\simeq& 5\times10^{-8} ~ \left( \frac{y_\nu}{10^{-12}} \right)^{-1}
	\left(\frac{m_{3/2}}{1{\mathrm{TeV}}}\right)
	\left(\frac{m_{\tilde{\nu}_R}}{1{\mathrm{TeV}}}\right)^{-1}
	~~~~~~~~~~~~~\nonumber\\
	&& \times\left(\frac{M_{I_L}}{10^{14}{\mathrm{GeV}}}\right)^{2}
	\left(\frac{M_{I_R}}{10^{16}{\mathrm{GeV}}}\right)^{-1}
	\left(\frac{T_R}{10^5{\mathrm{GeV}}}\right)|A|\delta_{\mathrm{eff}}
\end{eqnarray}
for $M_{I_L} \ll M_{I_R}$.
We used $H_{\mathrm{quar}} = 0.3\times{\mathrm{max}}\{{y_\nu M_{I_L}, y_\nu M_{I_R}}\}$
from numerical calculations.
Here, an uncertainty exists in the these three equations
as discussed in the previous section.
In the case that thermal-log terms
destabilize the AD-fields,
the asymmetry is estimated as
\begin{eqnarray}
	\label{eq:analytic_baryon-to-entropy_therlog}
	\frac{n_{B}}{s} &\simeq& 
	4\times10^{-13} ~ 
	\sum a_{\mathrm{th}}
	\left( \frac{y_\nu}{10^{-12}} \right)
	\left(\frac{m_{3/2}}{1{\mathrm{TeV}}}\right)
	~~~~~~~~~~~~~~~\nonumber\\
	&& \times \left(\frac{M_{I_L}}{10^{14}{\mathrm{GeV}}}\right)^8
	\left(\frac{M_{I_R}}{10^{14}{\mathrm{GeV}}}\right)
	\left(\frac{T_R}{10^9{\mathrm{GeV}}}\right)^{-5}
	|A| \delta_{\mathrm{eff}} Q_{\mathrm{th}}
\end{eqnarray}
where we used 
$H_{\mathrm{log}} = 0.2\times \sum a_{\mathrm{th}} \alpha_s^2 T_R^2 M_{\mathrm{Pl}}/M_{I_L}^2$
from numerical calculations.
Finally, in the case that thermal-mass terms
destabilize the AD-fields,
\begin{eqnarray}
	\label{eq:analytic_baryon-to-entropy_thermass}
	\frac{n_{B}}{s} &\simeq& 
	3\times10^{-10} ~ 
	\left( \frac{c_k f_k^{2}}{1.1\times10^{-10}} \right)^{-2}
	\left( \frac{y_\nu}{10^{-12}} \right)
	\left(\frac{m_{3/2}}{1{\mathrm{TeV}}}\right)
	~~~~~~~~~~~~~~~\nonumber\\
	&& \times \left(\frac{M_{I_L}}{10^{14}{\mathrm{GeV}}}\right)^2
	\left(\frac{M_{I_R}}{10^{14}{\mathrm{GeV}}}\right)
	\left(\frac{T_R}{10^8{\mathrm{GeV}}}\right)^{-1}
	 |A| \delta_{\mathrm{eff}} Q_{\mathrm{th}}
\end{eqnarray}
where $k$ indicates the particle giving the largest thermal-mass.
Here we used
$H_{\mathrm{th}} = 0.2\times {\mathrm{max}}
\{ H_{{\mathrm{th}},k} \theta (H_{{\mathrm{th}},k} - H_{{\mathrm {thr}},k})\}$
from numerical calculations.
In most cases we are interested in, 
only up (s)quarks contributes and 
$n_B/s \simeq 3\times10^{-10} |A| \delta_{\mathrm{eff}} Q_{\mathrm{th}}$
for typical parameters shown above.

\begin{figure}[t]
	\begin{center}
		\scalebox{.7}{\includegraphics{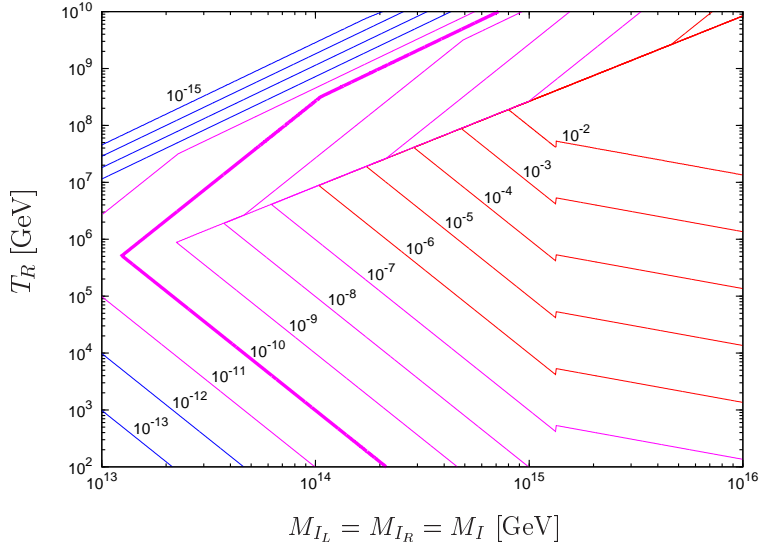}}
		\caption{\small The baryon-to-entropy ratio for $M_{I_L}=M_{I_R}=M_I$ is shown.
		A bold line indicates $n_B/s \sim 10^{-10}$.
		Parameters are the same as for Fig.~\ref{fig:destabilize}.
		}
		\label{fig:baryon-to-entropy_ratio}
	\end{center}
\end{figure}

The baryon-to-entropy ratio dependent on $M_I$ and $T_R$ are shown
in Fig.~\ref{fig:baryon-to-entropy_ratio}.
Here, $M_{I_R}=M_{I_L}=M_I$ is assumed.
Other parameters are the same as for Fig.~\ref{fig:destabilize}.
For simplicity, we assume $Q_{\mathrm{th}}=1$.
A bold line indicates $n_B/s = 10^{-10}$.
Compared with Fig.~\ref{fig:destabilize},
we can see which terms destabilize $\phi$.
If we assume some dilution process such as thermal inflation \cite{thermalinflation},
whole region $n_B/s >10^{-10}$ is allowed.
Therefore, sufficient baryon number asymmetry can be produced 
in the large parameter region $M_I\gtrsim10^{13}{\mathrm{GeV}}$.
To avoid gravitino overproduction, the reheating temperature should be less than
$10^9 {\mathrm{GeV}}$ \cite{gravitino}.
In the region $M_I \sim 10^{13}-10^{15} {\mathrm{GeV}}$, 
the baryon asymmetry can be explained without entropy production
even for high reheating temperature. 
Moreover, the resultant baryon number asymmetry is reduced 
if we take smaller $y_{\nu}$ as discussed later.
In those region, $n_B/s\sim10^{-10}$ can be realized for higher $T_R$
without entropy production.

\begin{figure}[p] 
	\begin{center}			
		\scalebox{.7}{\includegraphics{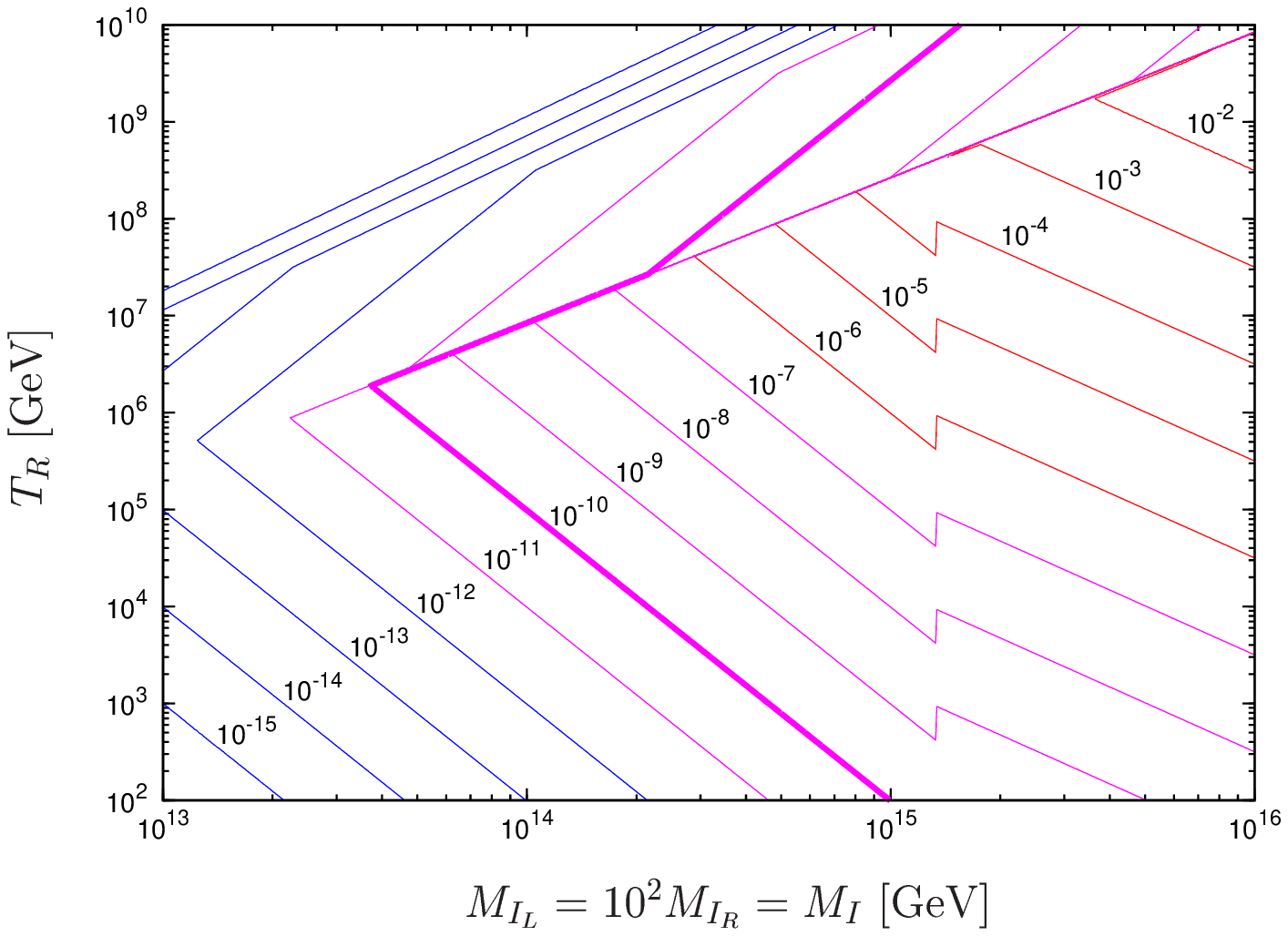}}
		\caption{\small The baryon-to-entropy ratio
		for $M_{I_L}=10^2M_{I_R}=M_I$ is shown.
		A bold line indicates $n_B/s \sim 10^{-10}$.
		Parameters are the same as for Fig.~\ref{fig:destabilize}.
		}
		\label{fig:baryon-to-entropy_ratio_R<L}
		\vspace{5mm}
		\scalebox{.7}{\includegraphics{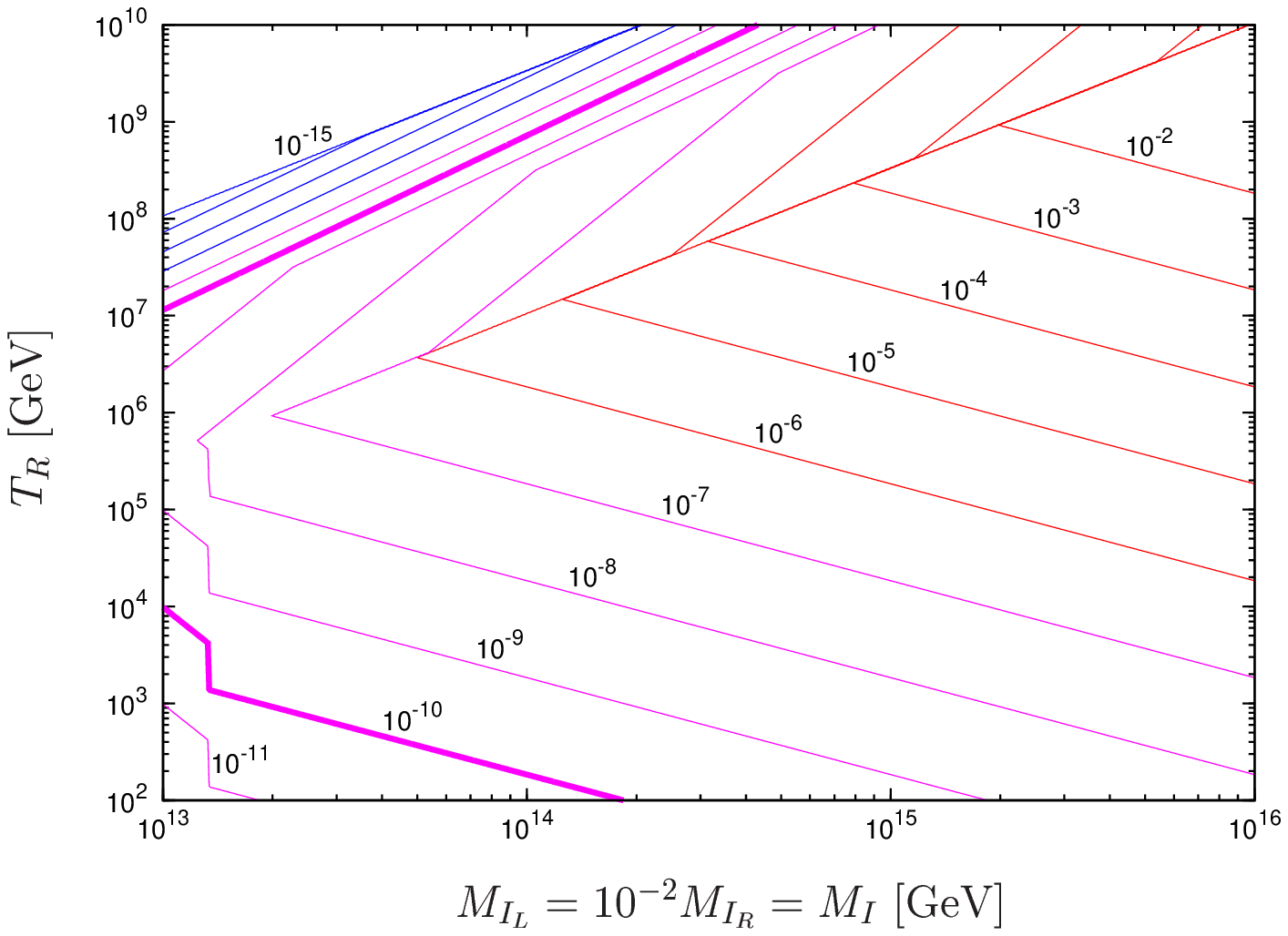}}
		\caption{\small The baryon-to-entropy ratio
		for $M_{I_L}=10^{-2}M_{I_R}=M_I$ is shown.
		A bold line indicates $n_B/s \sim 10^{-10}$.
		Parameters are the same as for Fig.~\ref{fig:destabilize}.
		}
		\label{fig:baryon-to-entropy_ratio_R>L}
	\end{center}
\end{figure}

If $M_{I_R} \neq M_{I_L}$,
the baryon-to-entropy ratio is modified by the factor $M_{I_L}/M_{I_R}$.
However, for destabilization by quartic terms,
the result can not be expressed in a simple form.
We also show the baryon-to-entropy ratio
in Fig.~\ref{fig:baryon-to-entropy_ratio_R<L} for $M_{I_L} > M_{I_R}$ and
in Fig.~\ref{fig:baryon-to-entropy_ratio_R>L} for $M_{I_L} < M_{I_R}$.
In these figures, parameters are the same as for Fig.~\ref{fig:destabilize},
except for $M_{I_L}$ and $M_{I_R}$.
In Fig.~\ref{fig:baryon-to-entropy_ratio_R<L},
the baryon number asymmetry is small compared with Fig.~\ref{fig:baryon-to-entropy_ratio}.
This is because we take smaller initial value of $\tilde{\nu}_R$.
On the other hand, in Fig.~\ref{fig:baryon-to-entropy_ratio_R>L},
the asymmetry is larger than that of Fig.~\ref{fig:baryon-to-entropy_ratio},
since larger initial value of $\tilde{\nu}_R$ is taken.

\begin{figure}[t]
	\begin{center}					
		\scalebox{.7}{\includegraphics{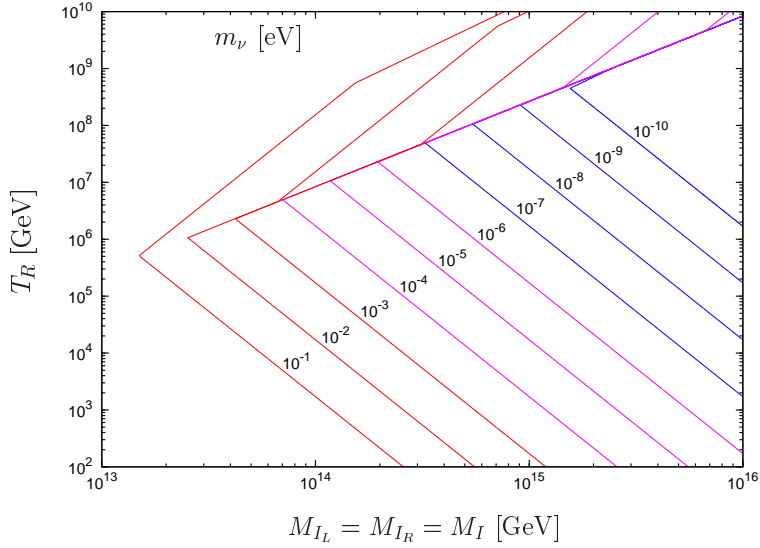}}
		\caption{\small The region giving $n_B/s\sim10^{-10}$ is shown
		dependent on the neutrino mass for $M_{I_L}=M_{I_R}=M_I$.
		Parameters are the same as for Fig.~\ref{fig:destabilize}.
		}
		\label{fig:baryon-to-entropy_ratio_var-ynu}
	\end{center}
\end{figure}

In Fig.~\ref{fig:baryon-to-entropy_ratio_var-ynu},
the region giving $n_B/s =10^{-10}$ is shown
dependent on the neutrino mass for $M_{I_L}=M_{I_R}=M_I$.
Parameters are the same as for Fig.~\ref{fig:destabilize}
except for $y_\nu$.
Note that $m_\nu = y_\nu \langle H_u \rangle$.
According to this result,
the amount of the baryon number asymmetry becomes smaller for smaller $m_{\nu}$.
This also means, inversely, $n_B/s\sim10^{-10}$ can be realized for smaller $m_\nu$
without dilution process.
This may be favored,
since high reheating temperature consistent with various inflation models
is allowed.

\section{Dark matter}

The number of right-handed sneutrino in condensate 
is bounded below by the baryon number asymmetry,
\begin{eqnarray}
	\label{eq:RH-sneutrino_number}
	n_{\tilde{\nu}_R} \gtrsim \frac{23}{8}B.
\end{eqnarray}
Assuming $R$-parity conservation,
the LSP produced by the decay of $\tilde{\nu}_R$, or $\tilde{\nu}_R$ itself,
remains as dark matter,
since the right-handed sneutrino decays after the freeze-out of the LSP at
$T \sim 5 - 50$ GeV,
unless $m_{\tilde{\nu}_R}$ is finely tuned to the left-handed sneutrino mass.
The abundance of dark matter originated in $\tilde{\nu}_R$
requires that the LSP mass is less than 1 GeV
for avoiding overclose of the universe.
However, very light LSP mass is disfavored in gravity mediated SUSY breaking models.
Hence, dark matter from $\tilde{\nu}_R$
is likely to overclose the universe.

This problem can not be avoided by late time entropy production
such as thermal inflation \cite{thermalinflation}.
Entropy production dilutes dark matter and baryon number simultaneously.
Hence, an appropriate amount of the baryon number asymmetry should be left in the universe
after entropy production.
Therefore, Eq.~(\ref{eq:RH-sneutrino_number}) is not changed
by entropy production.

If we assume SU(2)$_R$ gauge interaction broken at an intermediate scale $v_R$,
this problem can be solved.
By this interactions with $v_R \lesssim 10^6 - 10^8$ GeV,
$\tilde{\nu}_R$ can decay into right-handed particles
before the LSP freeze-out and after the electroweak phase transition.
One may worry that the SU(2)$_R$ broken at $v_R$
spoils the successful baryogenesis with $M_{I_R} \gtrsim v_R$.
However, this is not the case.
The direction $\phi = \tilde{\nu}_R$ is flat for the SU(2)$_R$ gauge.
In this case, Affleck-Dine mechanism is realized 
by the dynamics of only one complex scalar field 
corresponding to this flat direction.
The potential of this scalar field is almost the same as $\phi$,
therefore the difference is only that the left-right asymmetry becomes fixed
when this scalar field begins oscillation.
Hence, our estimate for $M_{I_L} = M_{I_R} = M_I$ are correct
even for the existence of SU(2)$_R$ gauge interaction.

\section{Summary}
\label{summary}

We have investigated the baryogenesis scenario
via left-right asymmetry generation by the Affleck-Dine mechanism
in SUSY standard models with Dirac neutrinos.
Only the left-handed lepton asymmetry is transferred to the baryon number
asymmetry via the sphaleron process,
while the right-handed lepton asymmetry remains in the right-handed sneutrinos
due to their tiny Yukawa couplings.
We explicitly include intermediate scale physics
to stabilize the right-handed sneutrino direction.
The evolutions of the AD-fields have been traced with thermal effects in detail.
Hence, the baryon number asymmetry can be estimated in broad parameter region.
As a result, it is pointed out that higher reheating temperature is allowed
without late time entropy production,
if the oscillation of $\phi$ is induced by thermal effects
or the lightest neutrino mass is small,
contrary to the previous work by Abel and Page.
We have also pointed out that
dark matter overproduction by the right-handed sneutrino decay
can be avoided by SU(2)$_R$ gauge interaction.

\section*{Acknowledgements}

The work of MS was supported in part by a
Grant-in-Aid of the Ministry of Education,
Culture, Sports, Science, and Technology,
Government of Japan, No. 18840011.


\end{document}